\documentclass[11pt]{article}
\usepackage{amssymb}
\usepackage{amsmath}
\usepackage{amsfonts}
\usepackage[singlespacing]{setspace}
\usepackage{geometry}
\usepackage{paralist}
\usepackage{hyperref}

\hypersetup{
  colorlinks=true,
  citecolor=blue,
  linkcolor=blue,
  filecolor=magenta,
  urlcolor=cyan,
}

\makeatletter
\renewcommand\@biblabel[1]{\textbullet}
\makeatother

\newtheorem{theorem}{Theorem}

\newtheorem{algorithm}[theorem]{Algorithm}

\newtheorem{definition}[theorem]{Definition}

\newtheorem{proposition}[theorem]{Proposition}

\newenvironment{proof}[1][Proof]{\noindent\textbf{#1.} }{\ \rule{0.5em}{0.5em}}
\normalsize
\begin{document}

\title{Tail approximations for sums of dependent regularly varying random
variables under Archimedean copula models}
\author{H\'{e}l\`{e}ne COSSETTE$^{\ast }$, Etienne MARCEAU\thanks{%
Universit\'{e} Laval, \'{E}cole d'Actuariat, 2425, rue de l'Agriculture,
Pavillon Paul-Comtois, local 4177, Qu\'{e}bec (Qu\'{e}bec) G1V 0A6, Canada,
Canada}, Quang Huy NGUYEN$^{\dag }$\ ~ \and and\ Christian Y. ROBERT\thanks{%
Universit\'{e} de Lyon, Universit\'{e} Lyon 1, Institut de Science Financi%
\`{e}re et d'Assurances, 50 Avenue Tony Garnier, F-69007 Lyon, France}}
\maketitle

\begin{abstract}
In this paper, we compare two numerical methods for approximating the
probability that the sum of dependent regularly varying random variables
exceeds a high threshold under Archimedean copula models. The first method
is based on conditional Monte Carlo. We present four estimators and show
that most of them have bounded relative errors. The second method is based
on analytical expressions of the multivariate survival or cumulative
distribution functions of the regularly varying random variables and
provides sharp and deterministic bounds of the probability of exceedance. We
discuss implementation issues and illustrate the accuracy of both procedures
through numerical studies.
\end{abstract}


\newpage

\section{Introduction}

A well-known problem in applied probability is the evaluation of the
probability that the sum of $n$ random variables (rvs) exceeds a certain
level $s$. This problem finds its way in many areas of application such as
actuarial science, finance, quantitative risk management and, reliability.\
Different methods can be used to tackle this problem depending on the type
of distributions of the random variables and their interaction as well as
the values of $n$ and $s$.

In this paper, we consider the case of $n$ positive heavy-tailed random
variables that are linked through dependence structures based on Archimedean
copulas. An $n$-dimensional copula $C$ is a multivariate distribution on $%
\left[ 0,1\right] ^{n}$ with uniform margins. Following \cite{4Ling65},
any Archimedean copula can be simply written as%
\begin{equation}
C(u_{1},...,u_{n})=\Phi \left( \Phi ^{\leftarrow }(u_{1})+...+\Phi
^{\leftarrow }(u_{n})\right) ,\text{ }(u_{1},...,u_{n})\in \left[ 0,1\right]
^{n},  \label{Definition Archi cop}
\end{equation}%
where $\Phi $ is a non-increasing function referred to as the generator of
the copula $C$ and $\Phi ^{\leftarrow }$ is the generalized inverse function
of $\Phi $ defined as $\Phi ^{\leftarrow }(x)=\inf \{t\in \mathbb{R}%
^{+}:\Phi (t)\leq x\}$. The conditions under which a generator $\Phi $
defines a proper $n$-dimensional copula are given in detail in \cite{4MN09}.

Let $\mathbf{U}=\left( U_{1},...,U_{n}\right) $ be a random vector
distributed as an Archimedean copula $C$. For $i=1,\ldots ,n$, we assume
that the survival distribution function of the $i$-th rv of the sum is
regularly varying, i.e. it satifies $\bar{F}_{i}(x)=1-F_{i}(x)=x^{-\alpha
_{i}}l_{i}(x)$ where $\alpha _{i}>0$ and $l_{i}$ is a slowly varying
function at infinity. We assume without loss of generality that $\alpha
_{1}\leq \ldots \leq \alpha _{n}$. Throughout the paper we shall consider
two sums that are linked through the Archimedean copula $C$ in the following
way%
\begin{equation*}
S_{n}^{X}=\sum_{i=1}^{n}X_{i}\quad \text{and}\quad
S_{n}^{Y}=\sum_{i=1}^{n}Y_{i}\text{,}
\end{equation*}%
where
\begin{equation*}
\mathbf{X}=(X_{1},...,X_{n})\overset{d}{=}(F_{1}^{\leftarrow
}(U_{1}),...,F_{n}^{\leftarrow }(U_{n}))\quad \text{and}\quad \mathbf{Y}%
=(Y_{1},...,Y_{n})\overset{d}{=}\left( \overline{F}_{1}^{\leftarrow
}(U_{1}),...,\overline{F}_{n}^{\leftarrow }(U_{n})\right) ,
\end{equation*}%
with $F_{i}^{\leftarrow }(u_{i})=\inf \{t\in \mathbb{R}^{+}:F_{i}(t)\geq x\}$
since $F_{i}$ is a non-decreasing function. Note that the multivariate
cumulative distribution function of $\mathbf{X}$ is given by
\begin{equation*}
\Pr (X_{1}\leq x_{1},...,X_{n}\leq x_{n})=C(F_{1}(x_{1}),...,F_{n}(x_{n})),%
\text{ }(x_{1},...,x_{n})\in \mathbb{R}^{n},
\end{equation*}%
while the multivariate survival distribution function of $\mathbf{Y}$ is
given by
\begin{equation*}
\Pr (Y_{1}>y_{1},...,Y_{n}>y_{n})=C\left( \overline{F}_{1}(y_{1}),...,%
\overline{F}_{n}(y_{n})\right) ,\text{ }(y_{1},...,y_{n})\in \mathbb{R}^{n}.
\end{equation*}%
The Archimedean copula $C$ is referred to as the copula of $\mathbf{X}$ and
as the survival copula of $\mathbf{Y}$.

To approximate the probabilities $z_{X}(s)=\Pr \left( S_{n}^{X}>s\right) $
and $z_{Y}(s)=\Pr \left( S_{n}^{Y}>s\right) $ when $s$ is large, one could
use functions that are asymptotically equivalent to these probabilities.
However there are few results concerning the asymptotic behaviours of these
probabilities. Actually it strongly depends on the tails of the Archimedean
copulas (see \cite{4CS09} for a fine analysis of the
several types of tails based on characteristics of the Archimedean
generator) and strong assumptions have to hold to characterize these
behaviours. For example, if the upper-tails of the Archimedean copula $C$
are independent, i.e $\lim_{u\rightarrow 1}\Pr (U_{i}>u|U_{j}>u)=0$ for all $%
i\neq j$ and if there exist $n-1$ non-negative constants $c_{2},...,c_{n}$
such that $c_{i}=\lim_{s\rightarrow \infty }\bar{F}_{i}(s)/\bar{F}_{1}(s)$,
then it can be shown that
\begin{equation*}
\lim\limits_{s\rightarrow \infty }\frac{\Pr (S_{n}^{X}>s)}{\overline{F}%
_{1}(s)}=1+\sum_{i=2}^{n}c_{i}.
\end{equation*}%
If the lower-tails of the Archimedean copula $C$ are rather independent,
i.e. $\lim_{u\rightarrow 0}\Pr (U_{i}<u|U_{j}<u)=0$ for all $i\neq j$, then
\begin{equation*}
\lim\limits_{s\rightarrow \infty }\frac{\Pr (S_{n}^{Y}>s)}{\overline{F}%
_{1}(s)}=1+\sum_{i=2}^{n}c_{i}
\end{equation*}%
(see e.g. \cite{4JM06} or \cite{4YY12}). \cite{4SL10} studied the asymptotic behaviours of $z_{X}(s)$ and $%
z_{Y}(s)$ under the assumption of identically distributed marginals and
specific upper or lower-tail dependence. Let $\beta >0$ and $l_{\Phi }$ be a
slowly varying function at infinity. If $1-\Phi (x)=x^{\beta }l_{\Phi
}\left( x^{-1}\right) $, they proved that%
\begin{equation*}
\lim\limits_{s\rightarrow \infty }\frac{\Pr (S_{n}^{X}>s)}{\Pr (X_{1}>s)}%
=q_{n}^{C}(\alpha ,\beta ),
\end{equation*}%
where $\alpha $ denotes the common tail index and
\begin{equation*}
q_{n}^{C}(\alpha ,\beta )=\int_{\sum_{i=1}^{n}v_{i}^{-1/\alpha }>1}\frac{%
\partial ^{n}}{\partial v_{1}...\partial v_{n}}\sum_{1\leq
i_{1},...,i_{j}\leq n}\left( (-1)^{j-1}(v_{i_{1}}^{1/\beta
}+...+v_{i_{j}}^{1/\beta })^{\beta }\right) \mathrm{d}v_{1}...\mathrm{d}%
v_{n}.
\end{equation*}%
If the generator rather satisfies $\Phi (x)=x^{-\beta }l_{\Phi }\left(
x\right) $, they proved that
\begin{equation*}
\lim\limits_{s\rightarrow \infty }\frac{\Pr (S_{n}^{Y}>s)}{\Pr (Y_{1}>s)}%
=q_{n}^{D}(\alpha ,\beta ),
\end{equation*}%
where
\begin{equation*}
q_{n}^{D}(\alpha ,\beta )=\int_{\sum_{i=1}^{n}v_{i}^{-1}>1}\frac{\partial
^{n}}{\partial v_{1}...\partial v_{n}}\left( v_{1}^{-\alpha /\beta
}+...+v_{n}^{-\alpha /\beta }\right) ^{-\beta }\mathrm{d}v_{1}...\mathrm{d}%
v_{n},
\end{equation*}%
(see also \cite{4W03}). Although $q_{n}^{C}(\alpha ,\beta )$
and $q_{n}^{D}(\alpha ,\beta )$ are known, they do not have closed-form
expressions and they can not be easily computed.

In this paper, we aim to provide two numerical methods for approximating $%
z_{X}(s)$ and $z_{Y}(s)$ for different values of $n$ and $s$ and choices of
parameters and functions: $((\alpha _{1},l_{1}),\ldots ,(\alpha _{n},l_{n}))$
and $\Phi $.

Our first method is based on conditional Monte Carlo and is ideally suited
when $s$ and/or $n$ are large. The classical Monte Carlo method is easy to
implement and can be applied in complex situations such as high dimensional
calculations. However, it is well known that it is inadequate for small
probability simulation since the relative errors (variation coefficients)
are too large. \cite{4AG06} introduced relative error as
a measure of efficiency of an estimator and several definitions of efficient
estimators. An unbiased estimator $Z(s)$ of the probability $z(s)$, with
relative error $e(Z(s))=\sqrt{\mathbb{E[}Z^{2}(s)]}/z(s)$, is called (i) a
logarithmically efficient estimator if $\lim \sup_{s\rightarrow \infty
}e(Z(s))\ [z(s)]^{\epsilon }=0$ for all $\epsilon >0$; (ii) an estimator
with bounded relative error if $\lim \sup_{s\rightarrow \infty
}e(Z(s))<\infty $; (iii) an estimator with vanishing relative error $\lim
\sup_{s\rightarrow \infty }e(Z(s))=0$.

For sums of independent random variables, the most widely used alternatives
to crude Monte Carlo computation of rare-event probabilities are conditional
Monte Carlo and importance sampling. \cite{4AB97}
propose a logarithmically efficient algorithm based on conditional Monte
Carlo simulation using order statistics. 

\cite{4BS01}
use importance sampling to simulate ruin probabilities for subexponential
claims and \cite{4KYT6} use importance sampling based
on hazard rate twisting to simulate heavy-tailed processes. \cite{4AK06} propose two algorithms which use importance sampling and
conditional Monte carlo and study their efficiency in the Pareto and Weibull
case.

Estimating tail distribution of the sums of dependent random variables via
simulation requires a specific expression for the dependence structure or a
closed form expression for the conditional distribution functions, the case
of elliptic distributions is an example. For an elliptic dependence
structure, \cite{4BR11} proposed a conditional
Monte Carlo estimator for the tail distribution of the sum of log-elliptic
random variables and proved that it has a logarithmically efficient relative
error. The sum of the log-elliptic random variables was also estimated by \cite{4KH13} using the simulation method introduced
by \cite{4AK06} and favorable results are presented
especially in the multivariate lognormal case. \cite{4ABJR11}
and \cite{4BAL8} focus on the efficient estimation of sums
of correlated lognormals using importance sampling and conditional Monte
Carlo strategies. \cite{4CK10}, \cite{4CK11} use conditional
Monte Carlo notably in a credit risk setting under the t-copula model to
estimate rare-event probabilities.

In this paper, we introduce four different estimators of the probabilities $%
z_{X}(s)=\Pr \left( S_{n}^{X}>s\right) $ and $z_{Y}(s)=\Pr \left(
S_{n}^{Y}>s\right) $ using techniques of conditional Monte Carlo simulation.
The main idea to build our estimators is to first isolate the known
probabilities $\Pr (M_{n}^{X}>s)$ or $\Pr (M_{n}^{Y}>s)$ where $M_{n}^{A}$
correspond to the maximum element of a given vector $\mathbf{A}$ (because $%
\Pr (M_{n}^{X}>s)$ or $\Pr (M_{n}^{Y}>s)$ have closed-form expressions in
our framework), and then simulate conditionally on the values taken by these
maxima. Two effective simulation techniques of vectors of Archimedean copula
proposed in \cite{4BHC13} and \cite{4MN09} will be used. We show that most of our estimators have bounded
relative errors.

Our second method is based on analytical expressions of the survival
multivariate distribution function and provides sharp, deterministic and
numerical bounds of the probabilities using the same ideas as developed in \cite{4CCMM14}. This method performs very well for cases
when $n$ is relatively small and effectively completes the conditional Monte
Carlo method.

The outline of the paper is as follows. In Section 2, we present the two
simulation techniques related to Archimedean copulas which are later used to
develop the proposed estimators. These estimators are introduced, described
and discussed in Section 3. Some results on their asymptotic efficiency are
also given. Section 4 explains how to derive and compute the numerical
bounds following the approach proposed in \cite{4CCMM14}.
Section 5 illustrates the accuracy of both methods through a numerical study
and discusses implementation issues.

\section{Simulation and conditional simulation with Archimedean copulas}

\label{section42} The classical simulation method for a dependent vector
relies on conditional distributions. Consider a random vector $\mathbf{U}$
with density function $c(u_{1},...,u_{n})$ which can be decomposed as the
product of conditional densities
\begin{equation*}
c(u_{1},...,u_{n})=c_{n|n-1,\ldots
,1}(u_{n}|u_{n-1},...,u_{1})...c_{2|1}(u_{2}|u_{1})c_{1}(u_{1}).
\end{equation*}%
The classical procedure of simulating vector such a vector $\mathbf{U}$ is
then: simulate $u_{1}$ based on $c_{1}(u_{1})$, simulate $U_{2}$ based on $%
c_{2|1}(u_{2}|u_{1})$, ..., simulate $U_{n}$ based on $%
c_{n|n-1,...,1}(u_{n}|u_{n-1},...,u_{1})$. Hence, the realization of vector $%
\mathbf{U}$ is created by calculating $(n-1)$ times the inverses of
conditional distribution functions. However it can be difficult and take
quite an amount of time when the distribution function of $\mathbf{U}$ is an
Archimedean copula.

Another method for an Archimedean copula could be to consider the mixed
exponential or frailty representation often used to model dependent
lifetimes and discussed in, notably, \cite{4MO88}, \cite{4Mc08} and \cite{Hofert2008}. In this case, the Archimedean
generator $\Phi $ is the Laplace-Stieltjes transform of a non-negative
random variable. Such a method thus requires to invert the Laplace-Stieltjes
transform $\Phi $ which can not always be evaluated explicitly.

To circumvent these problems, one can resort to two effective simulation
techniques proposed in \cite{4BHC13} and \cite{4MN09}. More precisely, \cite{4BHC13} use the
Kendall distribution function while \cite{4MN09}
suggest a simulation method which relies on $\ell _{1}$-norm symmetric
distributions.

\subsection{Brechmann, Hendrich and Czado's approach\label{section421}}

Arguing that the classical method does not work due to the problem of
calculating the inverse functions of conditional distributions
\begin{equation*}
C_{j|j-1,...,1}(u_{j}|u_{j-1},...,u_{1})=\Pr (U_{j}\leq
u_{j}|u_{j-1},...,u_{1}),
\end{equation*}%
\cite{4Br14} provides an algorithm to simulate Archimedean copulas
using an intermediate variable $Z$ whose distribution function is known as
the Kendall distribution function (see \cite{4BGGR96}). This
conditional inverse simulation method eliminates the problems encountered
with the numerical calculations of the inverse functions of $C_{j|j-1,...,1}$%
. We restate below two propositions which will prove useful for the
understanding of the algorithm proposed by \cite{4BHC13}.

\begin{proposition}[Barbe et al., 1996]
\label{Barbe1996} Let $\mathbf{U}$ be distributed as the Archimedean copula $%
C$ $\ $with generator $\Phi $ and let the random variable $Z$ be defined as $%
Z=C(\mathbf{U})$. Then, the density function of $Z$ is defined in terms of
the generator $\Phi $ as%
\begin{equation*}
f_{Z}(z)=\frac{(-1)^{n-1}}{(n-1)!}\ \left( \Phi ^{\leftarrow }(z)\right)
^{n-1}\ (\Phi ^{\leftarrow })^{(1)}(z)\ \Phi ^{(n)}\left( \Phi ^{\leftarrow
}(z)\right) .
\end{equation*}
\end{proposition}

\begin{proposition}[Brechmann, 2014]
\label{Brechmann2013} Let $\mathbf{U}$ be distributed as the Archimedean
copula $C$ with generator $\Phi $ and let the random variable $Z$ be defined
as $Z=C(\mathbf{U})$. Then, the conditional distribution of $%
U_{j}|Z,U_{j-1},...,U_{1}$ for $j=1,...,n$ is
\begin{equation*}
F_{U_{j}|Z,U_{j-1},...,U_{1}}(u_{j}|z,u_{j-1},...,u_{1})=\left( 1-\frac{\Phi
^{\leftarrow }(u_{j})}{\Phi ^{\leftarrow }(z)-\sum_{k=1}^{j-1}\Phi
^{\leftarrow }(u_{k})}\right) ^{n-j}
\end{equation*}%
for $1>u_{j}>\Phi \left( \Phi ^{\leftarrow }(z)-\sum_{k=1}^{j-1}\Phi
^{\leftarrow }(u_{k})\right) $.
\end{proposition}

From these results, the inverse function of the conditional distribution
function
\begin{equation*}
F_{U_{j}|Z,U_{j-1},...,U_{1}}(u_{j}|z,u_{j-1},...,u_{1})
\end{equation*}%
can be calculated with an explicit formula. Indeed, if we have $%
z,u_{j-1},...,u_{1}$ as realizations of $Z,U_{j-1},...,U_{1}$ respectively
and $v$ as a realization of a uniform random variable in $(0,1)$, a
realization of $U_{j}$ is obtained with%
\begin{equation*}
u_{j}=\Phi \left( (1-v^{1/(n-j)})\left( \Phi ^{\leftarrow
}(z)-\sum_{k=1}^{j-1}\Phi ^{\leftarrow }(u_{k})\right) \right) \text{.%
}
\end{equation*}%
The conditional distribution of $(Z\left\vert U_{1}\right. )$ is given in
the following proposition.

\begin{proposition}[Brechmann et al., 2013]
\label{Brechmann2013_fz|u1} Let $\mathbf{U}$ be distributed as the
Archimedean copula $C$ with generator $\Phi $ and let the random variable $Z$
be defined as $Z=C(\mathbf{U})$. Then, the conditional distribution $%
F_{Z|U_{1}}(z|u_{1})$ can be calculated by the Archimedean generator and its
derivatives as
\begin{equation}
F_{Z|U_{1}}(z|u_{1})=\left( \Phi ^{\leftarrow }\right)
^{(1)}(u_{1})\sum_{j=0}^{n-2}\frac{(-1)^{j}}{j!}\left( \Phi
^{\leftarrow }(z)-\Phi ^{\leftarrow }(u_{1})\right) ^{j}\Phi ^{(j+1)}(\Phi
^{\leftarrow }(z))\text{\textit{\ for }}z\in (0,u_{1}).
\label{Distribution Z|U1}
\end{equation}
\end{proposition}

Given the above propositions, the following algorithm is derived from \cite{4BHC13} to generate a random vector $%
(X_{1},...,X_{n}) $ from an $n$-dimensional Archimedean copula $C$ with
generator $\Phi $ or a random vector $(Y_{1},...,Y_{n})$ from an $n$%
-dimensional survival Archimedean copula $C$ with generator $\Phi $.

\begin{algorithm}
Brechmann et al. (2013)'s algorithm.

\begin{enumerate}
\item Generate a random variable $U_{1}$ uniformally distributed on (0,1).

\item Generate a random variable $\left( Z\left\vert U_{1}=u_{1}\right.
\right) $ from $\left( \ref{Distribution Z|U1}\right) $.

\item For $j=2,...,n:$ generate a random variable $\left( U_{j}\left\vert
Z=z,U_{1}=u_{1},...,U_{j-1}=u_{j-1}\right. \right) $ with%
\begin{equation*}
u_{j}=\Phi \left( (1-v^{1/(n-j)})\left( \Phi ^{\leftarrow
}(z)-\sum_{k=1}^{j-1}\Phi ^{\leftarrow }(u_{k})\right) \right) ,
\end{equation*}%
where $V$ has been generated as a random variable uniformally distributed on
(0,1).

\item Set $X_{i}=F_{X_{i}}^{\leftarrow }\left( U_{i}\right) $ or $Y_{i}=\bar{%
F}_{Y_{i}}^{\leftarrow }\left( U_{i}\right) $ for $i=1,...,n$.
\end{enumerate}
\end{algorithm}

As a consequence, it is easy to simulate a conditional Archimedean copula $%
U=(U_{1},...,U_{n}|U_{1}\in \lbrack a,b])$\ where $[a,b]\in (0,1)$\ by first
simulating $U_{1}$\ uniformly distributed on $[a,b]$ and then by following
Steps 2-4 of the previous algorithm.

\subsection{McNeil and Neslehov\`{a}'s approach\label{section422}}

\cite{4MN09} give the conditions under which a
generator $\Phi $ defines an $n$-dimensional copula by means of (\ref%
{Definition Archi cop}) and show that the close connection between
Archimedean copulas and $\ell _{1}$-norm symmetric distributions, introduced
by \cite{4Fang}, and that allows a new perspective and
understanding of Archimedean copulas. With such an insight, they are able to
consider cases where an Archimedean generator is not completely monotone or
equivalently is not equal to a Laplace transform of a non-negative random
variable (see \cite{4Kimber}).

\begin{theorem}[McNeil and Neslehov\`{a}, 2009]
\label{th1MN09} A real function $\Phi $: $[0,\infty )\rightarrow \lbrack
0,1] $ is the generator of an $n$-dimensional Archimedean copula if and only
if it is an $n$-monotone function on $[0,\infty )$ i.e it is differentiable
up to order $(n-2)$ and the derivatives satisfy $(-1)^{i}\Phi ^{(i)}(x)\geq
0,i=0,1,...,n-2$ for any $x$ in $[0,\infty )$ and further if $(-1)^{n-2}\Phi
^{(n-2)}$ is non-increasing and convex in $[0,\infty )$.
\end{theorem}

\begin{definition}[Fang and Fang, 1988]
A random vector $\mathbf{X}$ on $\mathbb{R}_{+}^{n}=\left[ 0,\infty \right)
^{n}$ follows an $\ell _{1}$-norm symmetric distribution if and only if
there exists a non-negative random variable $R$ independent of $\mathbf{W}$,
where $\mathbf{W}=(W_{1},...,W_{n})$ is a random vector distributed
uniformly on the unit simplex $\mathfrak{s}_{n}$,%
\begin{equation*}
\mathfrak{s}_{n}=\left\{ \mathbf{x}\in \mathbb{R}_{+}^{n}:\left\Vert \mathbf{%
x}\right\Vert _{1}=1\right\} ,
\end{equation*}%
so that $\mathbf{X}$ permits the stochastic representation
\begin{equation*}
\mathbf{X}\overset{d}{\mathbf{=}}R\mathbf{W}\text{.}
\end{equation*}%
The random variable $R$ is referred to as the radial part of $\mathbf{X}$
and its distribution as the radial distribution.
\end{definition}

The following theorem establishes the connection between $\ell _{1}$-norm
symmetric distributions and Archimedean copulas. More details and
interesting results and comments in that regard can be found in \cite{4MN09}.

\begin{theorem}[McNeil and Neslehov\`{a}, 2009]
\label{th2MN09} Let the random vector $\mathbf{U}\ $be distributed according
to an $n$-dimensional Archimedean copula $C$ with generator $\Phi $. Then, $%
\left( \Phi ^{-1}\left( U_{1}\right) ,...,\Phi ^{-1}\left( U_{n}\right)
\right) $ has an $\ell _{1}$-norm symmetric distribution with survival
copula $C$ and radial distribution $F_{R}$ given by
\begin{equation*}
F_{R}(x)=1-\sum_{j=0}^{n-2}(-1)^{j}\frac{x^{j}}{j!}\Phi
^{(j)}(x)-(-1)^{n-1}\frac{x^{n-1}}{(n-1)!}\Phi ^{(n-1)+}(x)\text{, }x\in %
\left[ 0,\infty \right) .
\end{equation*}
\end{theorem}

This last theorem implies
\begin{equation*}
(U_{1},...,U_{n})\,\overset{d}{=}(\Phi (RW_{1}),...,\Phi (RW_{n}))
\end{equation*}%
for $R$ a positive random variable with distribution function $F_{R}$ and $%
\mathbf{W}$ a random vector uniformly distributed on the $n$-dimensional
unit simplex $\mathfrak{s}_{n}$. Hence, since the vector $(X_{1},...,X_{n})$
has marginal distribution functions $F_{1},...,F_{n}$ and $F\left(
x_{1},...,x_{n}\right) =C\left( F_{1}\left( x_{1}\right) ,...,F_{n}\left(
x_{n}\right) \right) $ where $C$ is an Archimedean copula with generator $%
\Phi $, then
\begin{equation*}
(X_{1},...,X_{n})\overset{d}{=}(F_{1}^{\leftarrow }(\Phi
(RW_{1})),...,F_{n}^{\leftarrow }(\Phi (RW_{n}))).
\end{equation*}%
Since $(Y_{1},...,Y_{n})$ has marginal distribution functions $%
F_{1},...,F_{n}$ with a dependence structure defined through an Archimedean
survival copula $C$ with generator $\Phi $, then
\begin{equation*}
(Y_{1},...,Y_{n})\,\overset{d}{=}\,(\overline{F}_{1}^{\leftarrow }(\Phi
(RW_{1})),...,\overline{F}_{n}^{\leftarrow }(\Phi (RW_{n})))\text{.}
\end{equation*}%
This representation leads to the following sampling algorithm.

\begin{algorithm}
McNeil and Neslehov\`{a}'s algorithm.

\begin{enumerate}
\item Generate a vector $(E_{1},...,E_{n})$ of $n$ iid exponential rvs with
parameter $1$. Calculate $W_{i}=E_{i}/\sum_{j=1}^{n}E_{j}$ such that $%
\mathbf{W}$ is uniformly distributed on the $n$-dimensional unit simplex $%
\mathfrak{s}_{n}$.

\item Generate a random variable $R$ with distribution $F_{R}$ (see Theorem %
\ref{th2MN09}).

\item Return $\mathbf{U}$ where $U_{i}=\Phi \left( RW_{i}\right) $ for $%
i=1,...,n$.

\item Set $X_{i}=F_{X_{i}}^{\leftarrow }\left( U_{i}\right) $ or $Y_{i}=\bar{%
F}_{Y_{i}}^{\leftarrow }\left( U_{i}\right) $ for $i=1,...,n$.
\end{enumerate}
\end{algorithm}

\section{Estimators of $z_{X}(s)$ and $z_{Y}(s)$\label{section43}}

In this section, we propose four different estimators of $z_{X}(s)$ and $%
z_{Y}(s)$. All estimators rely on a similar idea which is to decompose the
probability of interest into different components. The known components are
exactly evaluated and the other ones, which are our main concern, are
estimated by simulation.

To begin, we need to establish basic notations. Throughout, $\mathbf{X}_{-i}$
denotes the vector $\mathbf{X}=(X_{1},...,X_{n})$ with the $i$-th component $%
X_{i}$ removed and $M_{i}^{X}$ (or $M_{i}\{\mathbf{X}\}$) corresponds to the
$i^{th}$ element of the vector $\mathbf{X}$ after rearranging the elements
of $\mathbf{X}$ in a non-decreasing order. Obviously, $M_{1}^{X}$ and $%
M_{n}^{X}$ correspond to the minimum and maximum element of vector $\mathbf{X%
}$, respectively. The same convention holds for $\mathbf{Y}%
=(Y_{1},...,Y_{n}) $.

In what follows, we encounter frequently the evaluation of the probabilities
$\Pr (M_{n}^{X}>s)$ and $\Pr \left( M_{n}^{Y}>s\right) $ which rely on the
marginal distributions $F_{1}$,...,$F_{n}$ and the Archimedean generator $%
\Phi $. They are obtained as follows
\begin{equation*}
\Pr (M_{n}^{X}>s)=1-C\left( F_{1}(s),...,F_{n}(s)\right)
\end{equation*}%
and
\begin{equation*}
\Pr (M_{n}^{Y}>s)=1-\overline{C}\left( \overline{F}_{1}(s),...,\overline{F}%
_{n}(s)\right) ,
\end{equation*}%
with $\overline{C}\left( u_{1},...,u_{n}\right) =\Pr \left(
U_{1}>u_{1},...,U_{n}>u_{n}\right) $.

\subsection{First estimator}

The first estimator of $z_{X}(s)$ and $z_{Y}(s)$ that we propose is based on
the simulation technique used by \cite{4BHC13} to generate
sampled values of the conditional vector $\left( \mathbf{U}|U_{1}\right) \in
\lbrack a,b]$ (see Section \ref{section421}). The idea is to first isolate
the known probability $\Pr (M_{n}^{X}>s)$ and then condition on the value
taken by the maximum $M_{n}^{X}$.

Hence, we have
\begin{eqnarray*}
z_{X}(s) &=&\Pr (M_{n}^{X}>s)+\Pr (S_{n}^{X}>s,s/n<M_{n}^{X}\leq s) \\
&=&\Pr (M_{n}^{X}>s)+\sum_{i=1}^{n}\Pr (s/n<X_{i}\leq s)\ \Pr
(S_{n}^{X}>s,X_{i}=M_{n}^{X}|s/n<X_{i}\leq s).
\end{eqnarray*}%
Then, it leads to the following estimator $Z_{NR1}^{X}(s)$ for $z_{X}(s)$:%
\begin{equation*}
Z_{NR1}^{X}(s)=\Pr (M_{n}^{X}>s)+\sum_{i=1}^{n}\left( \overline{F}_{i}(s/n)-%
\overline{F}_{i}(s)\right) I_{\{S_{n}^{X_{i}}>s,X_{i}^{i}=M_{n}^{X_{i}}\}},
\end{equation*}%
where $S_{n}^{X_{i}}$, $X_{i}^{i}$ and $M_{n}^{X_{i}}$ correspond to the
conditionnal random variables $S_{n}^{X_{i}}$, $X_{i}$ and $M_{n}^{X_{i}}$
given $\left( s/n<X_{i}\leq s\right) $. The challenging problem here is the
simulation of the vector
\begin{equation*}
\mathbf{X}^{i}=(X_{1}^{i},...,X_{n}^{i})=(X_{1},...,X_{n}|s/n<X_{i}\leq s)
\end{equation*}%
with an Archimedean copula as dependence structure. Given that $%
X_{i}=F_{i}^{\leftarrow }(U_{i})$, this last vector $\mathbf{X}^{i}$ can be
viewed as%
\begin{eqnarray*}
(X_{1},...,X_{n}|s/n<X_{i}\leq s) &\,\overset{d}{=}\,&(F_{1}^{\leftarrow
}(U_{1}),...,F_{n}^{\leftarrow }(U_{n})|s/n<F_{i}^{\leftarrow }(U_{i})\leq s)
\\
&\,\overset{d}{=}\,&(F_{1}^{\leftarrow }(U_{1}^{i+}),...,F_{n}^{\leftarrow
}(U_{n}^{i+})),
\end{eqnarray*}%
where $(U_{1}^{i+},...,U_{n}^{i+})=(U_{1},...,U_{n}|F_{i}(s/n)<U_{i}\leq
F_{i}(s))$.

Similarly for the random vector $\mathbf{Y}$ with multivariate cumulative
distribution function based on the Archimedean survival copula $C$, we have%
\begin{eqnarray*}
(Y_{1},...,Y_{n}|s/n<Y_{i}\leq s) &\,\overset{d}{=}\,&(\overline{F}%
_{1}^{\leftarrow }(U_{1}),...,\overline{F}_{n}^{\leftarrow }(U_{n})|s/n<%
\overline{F}_{i}^{\leftarrow }(U_{i})\leq s) \\
&\,\overset{d}{=}\,&(\overline{F}_{1}^{\leftarrow }(U_{1}^{i-}),...,%
\overline{F}_{n}^{\leftarrow }(U_{n}^{i-})),
\end{eqnarray*}%
where $(U_{1}^{i-},...,U_{n}^{i-})=(U_{1},...,U_{n}|\overline{F}%
_{i}(s/n)>U_{i}\geq \overline{F}_{i}(s))$. Hence, the first estimator for $%
z_{Y}(s)$ is given by
\begin{equation*}
Z_{NR1}^{Y}(s)=\Pr (M_{n}^{Y}>s)+\sum_{i=1}^{n}\left( \overline{F}_{i}(s/n)-%
\overline{F}_{i}(s)\right) I_{\{S_{n}^{Y_{i}}>s,Y_{i}^{i}=M_{n}^{Y_{i}}\}}.
\end{equation*}

We are now in a position to propose the following algorithms to generate
realizations of both estimators $Z_{NR1}^{X}(s)$ and $Z_{NR1}^{Y}(s)$.

\begin{algorithm}
(Estimator $Z_{NR1}^{X}(s)$) To generate a realization of $Z_{NR1}^{X}(s)$,
proceed as follows:

\begin{enumerate}
\item \textit{For }$i=1,...,n$\textit{, independently simulate }$U_{i}^{i+}$%
\textit{\ uniformly distributed on }$(F_{i}(s/n),F_{i}(s))$\textit{.}

\item \textit{For each }$U_{i}^{i+}$\textit{\ in the first step, simulate }$%
Z $\textit{\ based on conditional distribution }$F_{Z|U_{i}^{+}}$\textit{\
and then simulate }$(U_{1}^{i+},...,U_{i-1}^{i+},U_{i+1}^{i+},U_{n}^{i+})$%
\textit{.}

\item \textit{For each }$j=1,...,n$,\textit{\ compute }$X_{j}^{i}=F_{j}^{%
\leftarrow }(U_{j}^{i+})$\textit{\ and return }$I_{%
\{S_{n}^{X_{i}}>s,X_{i}^{i}=M_{n}^{X_{i}}\}}$\textit{\ which takes value 0
or 1. }

\item \textit{Return }$Z_{NR1}^{X}(s)=\Pr (M_{n}^{X}>s)+\sum_{i=1}^{n}\left(
\overline{F}_{i}(s/n)-\overline{F}(s)\right)
I_{\{S_{n}^{X_{i}}>s,X_{i}^{i}=M_{n}^{X_{i}}\}}$\textbf{.}
\end{enumerate}
\end{algorithm}

\begin{algorithm}
(Estimator $Z_{NR1}^{Y}(s)$) To generate a realization of $Z_{NR1}^{Y}(s)$,
proceed as follows:

\begin{enumerate}
\item \textit{For} $i=1,...,n$\textit{, independently simulate }$U_{i}^{i-}$%
\textit{\ uniformly distributed on }$(\bar{F}_{i}(s),\bar{F}_{i}(s/n))$%
\textit{.}

\item \textit{For each }$U_{i}^{i-}$\textit{\ in the first step, simulate }$%
Z $\textit{\ based on conditional distribution }$F_{Z|U_{i}^{-}}$\textit{\
and then simulate }$(U_{1}^{i-},...,U_{i-1}^{i-},U_{i+1}^{i-},U_{n}^{i-})$%
\textit{.}

\item \textit{For each }$j=1,2,\ldots ,n$\textit{, compute }$Y_{j}^{i}=\bar{F%
}_{j}^{\leftarrow }(U_{j}^{i-})$\textit{\ and return }$I_{%
\{S_{n}^{Y_{i}}>s,Y_{i}^{i}=M_{n}^{Y_{i}}\}}$\textit{\ which takes value 0
or 1.}

\item \textit{Return }$Z_{NR1}^{Y}(s)=\Pr (M_{n}^{Y}>s)+\sum_{i=1}^{n}\left(
\overline{F}_{i}(s/n)-\overline{F}_{i}(s)\right)
I_{\{S_{n}^{Y_{i}}>s,Y_{i}^{i}=M_{n}^{Y_{i}}\}}$.
\end{enumerate}
\end{algorithm}

\begin{proposition}
\label{ZNR1} Estimators $Z_{NR1}^{X}(s)$ and $Z_{NR1}^{Y}(s)$ have bounded
relative errors.
\end{proposition}

\begin{proof}
See Appendix.
\end{proof}

It is important to note that the approach used here to estimate the sum of
regularly varying random variables leads to estimators with a bounded
relative error no matter the dependence structure between the random
variables.The key element is the simulation of the conditional random vector
$\left( \mathbf{X}|s/n<X_{i}\leq s\right) $. Unfortunately, the numerical
performance of $Z_{NR1}(s)$, as we will see in Section \ref{section44}, is
not as good as for the other estimators.

\subsection{Second estimator}

The construction of the second estimator is based on the stochastic
representation of an Archimedean copula proposed by \cite{4MN09}. As stated in Section \ref{section422}, for a multivariate
random vector $\mathbf{X}$ with underlying Archimedean copula $C$ with
generator $\Phi $, we have
\begin{equation*}
(X_{1},...,X_{n})\,\overset{d}{=}\,(F_{1}^{\leftarrow }(\Phi
(RW_{1})),...,F_{n}^{\leftarrow }(\Phi (RW_{n}))),
\end{equation*}%
where the distribution function of $R$ is given as in Theorem \ref{th2MN09}
and $\mathbf{W}$ is a random vector uniformly distributed on $\mathfrak{s}%
_{n}$. This representation of the random vector $\mathbf{X}$ permits to
write the probability of interest $z_{X}(s)$ as follows:
\begin{eqnarray*}
z_{X}(s) &=&\Pr (M_{n}^{X}>s)+\Pr (S_{n}^{X}>s,M_{n}^{X}\leq s) \\
&=&\Pr (M_{n}^{X}>s)+\Pr \left( \sum_{i=1}^{n}F_{i}^{\leftarrow
}(\Phi (RW_{i}))>s,M_{n}\left\{ F_{i}^{\leftarrow }(\Phi (RW_{i}))\right\}
\leq s\right) \\
&=&\Pr (M_{n}^{X}>s)+\Pr \left( \sum_{i=1}^{n}F_{i}^{\leftarrow
}\left( \Phi (RW_{i})\right) >s,R\geq M_{n}\left\{ \frac{\Phi ^{\leftarrow
}(F_{i}(s))}{W_{i}}\right\} \right) .
\end{eqnarray*}%
By conditioning on the random vector $\mathbf{W}$, we have
\begin{eqnarray*}
&&\Pr \left( \sum_{i=1}^{n}F_{i}^{\leftarrow }\left( \Phi
(RW_{i})\right) >s,R\geq M_{n}\left\{ \frac{\Phi ^{\leftarrow }(F_{i}(s))}{%
W_{i}}\right\} \right) \\
&=&E_{\mathbf{W}}\left[ \Pr \left( \left.
\sum_{i=1}^{n}F_{i}^{\leftarrow }\left( \Phi (RW_{i})\right) >s,R\geq
M_{n}\left\{ \frac{\Phi ^{\leftarrow }(F_{i}(s))}{W_{i}}\right\} \right\vert
\mathbf{W}\right) \right] .
\end{eqnarray*}

Then, we obtain the following second estimator of $z_{X}(s)$ in terms of the
known radial cumulative distribution function $F_{R}$ given by
\begin{eqnarray*}
Z_{NR2}^{X}(s) &=&\Pr (M_{n}^{X}>s)+\Pr \left( \left.
\sum_{i=1}^{n}F_{i}^{\leftarrow }\left( \Phi (RW_{i})\right) >s,R\geq
M_{n}\left\{ \frac{\Phi ^{\leftarrow }(F_{i}(s))}{W_{i}}\right\} \right\vert
\mathbf{W}\right) \\
&=&\Pr (M_{n}^{X}>s)+\left( F_{R}(U^{X}(\mathbf{W},s))-F_{R}(L^{X}(\mathbf{W}%
,s))\right) ,
\end{eqnarray*}%
where
\begin{equation}
U^{X}(\mathbf{W},s)=\sup \{r\in \mathbb{R}^{+}:\sum_{i=1}^{n}F_{i}^{%
\leftarrow }(\Phi (rW_{i}))\leq s\}  \label{UX}
\end{equation}%
and%
\begin{equation}
L^{X}(\mathbf{W},s)=M_{n}\left\{ \frac{\Phi ^{\leftarrow }(F_{i}(s))}{W_{i}}%
\right\} .  \label{LX}
\end{equation}

In a similar fashion for the random vector $\mathbf{Y}$ with an underlying
Archimedean survival copula, we obtain the estimator $Z_{NR2}^{Y}(s)$ for $%
z_{Y}(s)$ which is given by
\begin{equation*}
Z_{NR2}^{Y}(s)=\Pr (M_{n}^{Y}>s)+\left( F_{R}(U^{Y}(\mathbf{W}%
,s))-F_{R}(L^{Y}(\mathbf{W},s))\right) ,
\end{equation*}%
where
\begin{equation}
U^{Y}(\mathbf{W},s)=M_{1}\left\{ \frac{\Phi ^{\leftarrow }(\overline{F}%
_{i}(s))}{W_{i}}\right\}  \label{UY}
\end{equation}%
and%
\begin{equation}
L^{Y}(\mathbf{W},s)=\inf \{r\in \mathbb{R}^{+}:\sum_{i=1}^{n}%
\overline{F}_{i}^{\leftarrow }(\Phi (rW_{i}))\geq s\}.  \label{LY}
\end{equation}

Note that if the marginal distributions are continuous and strictly
increasing, then $U^{X}(\mathbf{W},s)$ and $L^{Y}(\mathbf{W},s)$ are the
unique roots of equations $\sum_{i=1}^{n}F_{i}^{\leftarrow }(\Phi
(xW_{i}))=s$ and $\sum_{i=1}^{n}\overline{F}_{i}^{\leftarrow }(\Phi
(xW_{i}))=s$ respectively.

The sampling algorithm to generate realizations of $Z_{NR2}^{X}(s)$ and $%
Z_{NR2}^{Y}(s)$ can be written down as outlined in the following.

\begin{algorithm}
(Estimator $Z_{NR2}^{X}(s)$) To generate a realization of $Z_{NR2}^{X}(s)$,
proceed as follows:

\begin{enumerate}
\item Let $(E_{1},...,E_{n})$ be $n$ iid exponential rvs with parameter $1$.
Calculate $W_{i}=E_{i}/\sum_{j=1}^{n}E_{j}$.

\item Evaluate numerically $U^{X}(\mathbf{W},s)$ from $\left( \ref{UX}%
\right) $ and $L^{X}(\mathbf{W},s)$ from $\left( \ref{LX}\right) $.

\item Calculate derivatives $\Phi ^{(j)}$ for $j=1,...,n-1$ and then the
radial distribution $F_{R}.$

\item Return $Z_{NR2}^{X}(s)=\Pr (M_{n}^{X}>s)+F_{R}(U^{X}(\mathbf{W}%
,s))-F_{R}(L^{X}(\mathbf{W},s)).$
\end{enumerate}
\end{algorithm}

\begin{algorithm}
(Estimator $Z_{NR2}^{Y}(s)$) To generate a realization of $Z_{NR2}^{Y}(s)$,
proceed as follows:

\begin{enumerate}
\item Let $(E_{1},...,E_{n})$ be $n$ iid exponential rvs with parameter $1$.
Calculate $W_{i}=E_{i}/\sum_{j=1}^{n}E_{j}$.

\item Evaluate numerically $U^{Y}(\mathbf{W},s)$ from $\left( \ref{UX}%
\right) $ and $L^{Y}(\mathbf{W},s)$ from $\left( \ref{LX}\right) $.

\item Calculate derivatives $\Phi ^{(j)}$ for $j=1,...,n-1$ and then the
radial distribution $F_{R}.$

\item Return $Z_{NR2}^{Y}(s)=\Pr (M_{n}^{Y}>s)+F_{R}(U^{Y}(\mathbf{W}%
,s))-F_{R}(L^{Y}(\mathbf{W},s))$.
\end{enumerate}
\end{algorithm}

In the following proposition, the accuracy of our estimator $Z_{NR2}^{Y}(s)$
is investigated under the assumption that the Archimedean generator is
regularly varying.

\begin{proposition}
\label{ZNR2} For the random vector $\mathbf{Y}$ with an underlying
Archimedean survival copula $C$ with generator $\Phi $ satisfying $\Phi
^{(n-2)}$ differentiable and $\Phi (x)=x^{-\beta }l_{\Phi }\left( x\right) $
with $\beta >0$, then $Z_{NR2}^{Y}(s)$ has a bounded relative error.
\end{proposition}

\begin{proof}
See Appendix.
\end{proof}

\subsection{Third estimator}

This section presents the third estimator for $z_{X}(s)$ which will show
better numerical performances than the two previous estimators in the
numerical study presented in a later section. The third estimator for $%
z_{Y}(s)$ is based on the same idea and is not discussed. Let us separate
the probability $z_{X}(s)$ into the components
\begin{equation*}
\Pr (S_{n}^{X}>s)=\Pr (M_{n}^{X}>s)+z_{1}^{X}(s)+z_{2}^{X}(s),
\end{equation*}%
where $z_{1}^{X}(s)=\Pr (S_{n}^{X}>s,M_{n-1}^{X}\leq \lambda s,M_{n}^{X}\leq
s)$, $z_{2}(s)=\Pr (S_{n}^{X}>s,M_{n-1}^{X}>\lambda s,M_{n}^{X}\leq s)$ and $%
\lambda $ is a positive quantity less than $1/n$. In $z_{1}^{X}(s)$, the
inequality $M_{n-1}^{X}\leq \lambda s$ implies that there is only one
variable taking a large value. Consequently, we estimate $z_{1}^{X}(s)$
conditionally on $\mathbf{X}_{-i}$ when $X_{i}=M_{n}^{X}$. In $z_{2}^{X}(s)$%
, there are at least two variables taking large values, so it is coherent if
we estimate $z_{2}^{X}(s)$\ conditionally on the uniform random vector $%
\mathbf{W}$\ defined on the unit simplex $\mathfrak{s}_{n}$.

\subsubsection{Estimators for $z_{1}^{X}(s)$ and $z_{1}^{Y}(s)$}

Let us develop the probability $z_{1}^{X}(s)$ as
\begin{eqnarray*}
\Pr (S_{n}^{X}>s,M_{n-1}^{X}\leq \lambda s,M_{n}^{X}<s)
&=&\sum_{i=1}^{n}\Pr (S_{n}^{X}>s,M_{n-1}^{X}\leq \lambda
s,M_{n}^{X}<s,X_{i}=M_{n}^{X}) \\
&=&\sum_{i=1}^{n}\Pr (S_{n}^{X}>s,\max \{\mathbf{X}_{-i}\}\leq
\lambda s,X_{i}<s,X_{i}=M_{n}^{X}).
\end{eqnarray*}%
By conditioning on $\mathbf{X}_{-i}$, we obtain the following estimator $%
Z_{NR3,1}^{[i]X}(s)$ for $\Pr (S_{n}^{X}>s,\max \{\mathbf{X}_{-i}\}\leq
\lambda s,X_{i}<s,X_{i}=M_{n}^{X}):$
\begin{equation*}
Z_{NR3,1}^{[i]X}(s)=\Psi \left( \mathbf{X}_{-i},s\right) ,
\end{equation*}%
where
\begin{equation*}
\Psi \left( \mathbf{x}_{-i},s\right) =I_{\{\max \{\mathbf{x}_{-i}\}\leq
\lambda s\}}\ \Pr \left( s>X_{i}^{\ast }>s-\sum_{j=1,j\neq
i}^{n}x_{j}\right)
\end{equation*}%
with $X_{i}^{\ast }=\left( X_{i}|\mathbf{X}_{-i}=\mathbf{x}_{-i}\right) $
and $i=1,...,n$. Note that, if $\max \{\mathbf{x}_{-i}\}<\lambda s$, then
\begin{equation*}
s-\sum_{j=1,j\neq i}^{n}x_{j}>(1-(n-1)\lambda )s>s/n>\lambda s\geq
\max \{\mathbf{x}_{-i}\}
\end{equation*}%
which is coherent with $X_{i}=M_{n}^{X}>\max \{\mathbf{X}_{-i}\}$. Estimator
$Z_{NR3,1}^{X}(s)$ for $z_{1}^{X}(s)$ is then defined by
\begin{equation*}
Z_{NR3,1}^{X}(s)=\sum_{i=1}^{n}Z_{NR3,1}^{[i]X}(s).
\end{equation*}%
Under the assumption of identically distributed random variables $%
X_{1},...,X_{n}$, the estimator $Z_{NR3,1}^{X}(s)$ coincides with Asmussen
and Kroese's estimator (see \cite{4AK06}).

To perform the calculations, we need the conditional distribution of $%
X_{i}^{\ast }=\left( X_{i}|\mathbf{X}_{-i}=\mathbf{x}_{-i}\right) $ for each
$i=1,...,n$ which is given by
\begin{equation*}
F_{X_{i}^{\ast }}(x_{i})=\frac{\Phi ^{(n-1)}\left( \sum_{j=1}^{n}\Phi
^{-1}(F_{j}(x_{j}))\right) }{\Phi ^{(n-1)}\left( \sum_{j=1,j\neq
i}^{n}\Phi ^{-1}(F_{j}(x_{j}))\right) }.
\end{equation*}%
The method is similar to obtain $Z_{NR3,1}^{Y}(s)$ except for the expression
of the distribution of $Y_{i}^{\ast }$ which is slightly more difficult to
derive.

\begin{proposition}
\label{ConSurAr} Let $\mathbf{Y}=(Y_{1},...,Y_{n})$ with multivariate
distribution defined with an Archimedean survival copula and marginals $%
F_{1},...,F_{n}$. The conditional cumulative distribution function of $%
Y_{i}^{\ast }=\left( Y_{i}|\mathbf{Y}_{-i}=\mathbf{y}_{-i}\right) $ is
\begin{equation*}
F_{Y_{i}^{\ast }}(y_{i})=\Pr (Y_{i}^{\ast }\leq y_{i})=1-\frac{\Phi
^{(n-1)}\left( \sum_{j=1}^{n}\Phi ^{-1}(\bar{F}_{j}(y_{j}))\right) }{%
\Phi ^{(n-1)}\left( \sum_{j=1,j\neq i}^{n}\Phi ^{-1}(\bar{F}%
_{j}(y_{j}))\right) }\text{.}
\end{equation*}
\end{proposition}

\begin{proof}
See Appendix.
\end{proof}

\subsubsection{Estimators for $z_{2}^{X}(s)$ and $z_{2}^{Y}(s)$}

Given \cite{4MN09}, we can write the probability $%
z_{2}^{X}(s)=\Pr (S_{n}^{X}>s,M_{n-1}^{X}>\lambda s,M_{n}^{X}\leq s)$ as
\begin{eqnarray*}
z_{2}^{X}(s) &=&\Pr \left( \sum_{i=1}^{n}F_{i}^{\leftarrow }\left(
\Phi (W_{i}R)\right) >s,M_{n-1}\{F_{i}^{\leftarrow }\left( \Phi
(W_{i}R)\right) \}>\lambda s,M_{n}\{F_{i}^{\leftarrow }\left( \Phi
(W_{i}R)\right) \}\leq s\right) \\
&=&\Pr \left( R<U^{X}(\mathbf{W},s),R<M_{n-1}\left\{ \frac{\Phi ^{\leftarrow
}\left( F_{i}(\lambda s)\right) }{W_{i}}\right\} ,R\geq L^{X}(\mathbf{W}%
,s)\right) .
\end{eqnarray*}%
Conditioning on $\mathbf{W}$, we have
\begin{equation*}
z_{2}^{X}(s)=\mathbb{E}_{\mathbf{W}}\left[ \Pr \left( \left. R<U^{X}(\mathbf{%
W},s),R<M_{n-1}\left\{ \frac{\Phi ^{\leftarrow }\left( F_{i}(\lambda
s)\right) }{W_{i}}\right\} ,R\geq L^{X}(\mathbf{W},s)\right\vert \mathbf{W}%
\right) \right] ,
\end{equation*}%
which leads to the estimator $Z_{NR3,2}^{X}(s)$ given by
\begin{eqnarray*}
Z_{NR3,2}^{X}(s) &=&F_{R}\left( U^{X}(\mathbf{W},s)\wedge M_{n-1}\left\{
\frac{\Phi ^{\leftarrow }\left( F_{i}(\lambda s)\right) }{W_{i}}\right\}
\right) -F_{R}\left( L^{X}(\mathbf{W},s)\right) \\
&=&F_{R}\left( U_{\lambda }^{X}(\mathbf{W},s)\right) -F_{R}\left( L^{X}(%
\mathbf{W},s)\right)
\end{eqnarray*}%
with $U^{X}(\mathbf{W}^{\left( j\right) },s)$, $L^{X}(\mathbf{W}^{\left(
j\right) },s)$ as in (\ref{UX}) and (\ref{LX}) respectively, and
\begin{equation}
U_{\lambda }^{X}(\mathbf{W}^{\left( j\right) },s)=U^{X}(\mathbf{W}^{\left(
j\right) },s)\wedge M_{n-1}\left\{ \frac{\Phi ^{\leftarrow }\left(
F_{i}(\lambda s)\right) }{W_{i}^{\left( j\right) }}\right\} .
\label{UlamdaX}
\end{equation}%
Similarly, under an Archimedean survival copula, we have
\begin{eqnarray*}
z_{2}^{Y}(s) &=&\Pr \left( \sum_{i=1}^{n}\overline{F}_{i}^{\leftarrow
}\left( \Phi (W_{i}R)\right) >s,M_{n-1}\{\bar{F}_{i}^{\leftarrow }\left(
\Phi (W_{i}R)\right) \}\geq \lambda s,M_{n}\{\bar{F}_{i}^{\leftarrow }\left(
\Phi (W_{i}R)\right) \}\leq s\right) \\
&=&\Pr (R>L^{Y}(\mathbf{W},s),R\geq M_{2}\left\{ \frac{\Phi ^{\leftarrow }(%
\bar{F}_{i}(\lambda s))}{W_{i}}\right\} ,R\leq U^{Y}(\mathbf{W},s)).
\end{eqnarray*}%
The estimator $Z_{NR3,2}^{Y}(s)$ is hence given by%
\begin{eqnarray*}
Z_{NR3,2}^{Y}(s) &=&F_{R}\left( U^{Y}(\mathbf{W},s)\right) -F_{R}\left(
L^{Y}(\mathbf{W},s)\vee M_{2}\left\{ \frac{\Phi ^{\leftarrow }(\overline{F}%
_{i}(\lambda s))}{W_{i}}\right\} \right) \\
&=&F_{R}\left( U^{Y}(\mathbf{W},s)\right) -F_{R}\left( L_{\lambda }^{Y}(%
\mathbf{W},s)\right)
\end{eqnarray*}%
with $U^{Y}(\mathbf{W},s)$, $L^{Y}(\mathbf{W},s)$ as given in (\ref{UY}), (%
\ref{LY}) respectively, and
\begin{equation}
L_{\lambda }^{Y}(\mathbf{W},s)=L^{Y}(\mathbf{W},s)\vee M_{2}\left\{ \frac{%
\Phi ^{\leftarrow }(\overline{F}_{i}(\lambda s))}{W_{i}}\right\} .
\label{LlamdaY}
\end{equation}

\subsubsection{Estimators for $z_{X}(s)$ and $z_{Y}(s)$}

The third estimators for $z_{X}(s)$ and $z_{Y}(s)$ are finally given by%
\begin{equation*}
Z_{NR3}^{X}(s)=\Pr (M_{n}^{X}>s)+Z_{NR3,1}^{X}(s)+Z_{NR3,2}^{X}(s)\quad
\text{and}\quad Z_{NR3}^{Y}(s)=\Pr
(M_{n}^{Y}>s)+Z_{NR3,1}^{Y}(s)+Z_{NR3,2}^{Y}(s).
\end{equation*}

\begin{algorithm}
To generate a realization of $Z_{NR3}^{X}(s)$, proceed as follows:

\begin{enumerate}
\item Let $(E_{1},...,E_{n})$ be $n$ iid exponential rvs with parameter 1.
Calculate $W_{i}=E_{i}/\sum_{j=1}^{n}E_{j}$.

\item Compute $U_{\lambda }^{X}(\mathbf{W},s)$ from (\ref{UlamdaX}) and $%
L^{X}(\mathbf{W},s)$ from (\ref{LX}).

\item For $i=1,...,n$, simulate vector $\mathbf{U}_{-i}$ following $(n-1)$
dimensional Archimedean copula of generator $\Phi .$

\begin{enumerate}
\item Evaluate $X_{j}=F_{j}^{\leftarrow }(U_{j})$ for $j\neq i$.

\item Evaluate $Z_{NR3,1}^{[i]X}(s)=I_{\{\max \{\mathbf{x}_{-i}\}<\lambda
s\}}\left( F_{X_{i}^{\ast }}(s)-F_{X_{i}^{\ast }}(s-sum(\mathbf{x}%
_{-i}))\right) .$

\item Return $Z_{NR3}^{X}(s)=\Pr (M_{n}^{X}>s)+F_{R}(U_{\lambda
}^{X}(W,s))-F_{R}(L^{X}(W,s))+\sum_{i=1}^{n}Z_{NR3,1}^{[i]X}(s).$
\end{enumerate}
\end{enumerate}
\end{algorithm}

\begin{algorithm}
To generate a realization of $Z_{NR3}^{Y}(s)$, proceed as follows:

\begin{enumerate}
\item Let $(E_{1},...,E_{n})$ be $n$ iid exponential rvs of parameter 1,
calculate $W_{i}=E_{i}/\sum_{j=1}^{n}E_{j}$.

\item Calculate $U^{Y}(\mathbf{W},s)$ from (\ref{UY}) and $L_{\lambda }^{Y}(%
\mathbf{W},s)$ from (\ref{LlamdaY})

\item For $i=1,...,n$, simulate vector $U_{-i}$ following $(n-1)$
dimensional Archimedean copula of generator $\Phi $ and then calculate $%
Y_{j}=\bar{F}_{j}^{\leftarrow }(U_{j})$ for $j\neq i$. After that, calculate
the value of $Z_{NR3,1}^{[i]Y}(s)=I_{\{\max \{\mathbf{y}_{-i}\}<\lambda
s\}}\left( F_{Y_{i}^{\ast }}(s)-F_{Y_{i}^{\ast }}(s-sum(\mathbf{y}%
_{-i}))\right) $

\item Return $Z_{NR3}^{Y}(s)=P(M_{n}^{Y}>s)+F_{R}(U^{Y}(W,s))-F_{R}(L_{%
\lambda }^{Y}(W,s))+\sum_{i=1}^{n}Z_{NR3,1}^{[i]Y}(s).$
\end{enumerate}
\end{algorithm}

Unfortunately, the relative errors of $Z_{NR3,1}^{X}(s)$ and $%
Z_{NR3,1}^{X}(s)$ are not bounded if no assumption is made on the
Archimedean generator. Consequently, the relative errors of $Z_{NR3}^{X}(s)$
and $Z_{NR3}^{Y}(s)$ will not be bounded either in general. However,
numerical performances of these estimators are better than $Z_{NR2}(s)$ in
some situations when parameter $\lambda $\ takes appropriate values.
Moreover, in almost all cases, $Z_{NR3}^{X}(s)$ and $Z_{NR3}^{Y}(s)$ perform
better than $Z_{NR1}^{X}(s)$ and $Z_{NR1}^{Y}(s)$ which we have proven to
have a bounded relative error.

However we are able to prove the following result.

\begin{proposition}
\label{ZNR3} The estimator $Z_{NR3,2}^{Y}(s)$ has bounded relative error.
\end{proposition}

\begin{proof}
See Appendix.
\end{proof}

\subsubsection{Fourth estimator}

We propose in this section a fourth and final estimator of $z_{X}(s)$ which
is derived in a similar fashion as $Z_{NR3}^{X}(s)$, meaning that we split $%
\Pr (S_{n}^{X}>s,M_{n}^{X}\leq s)$ into two parts. Note that the estimator
under an Archimedean survival copula structure, denoted by $Z_{NR4}^{Y}(s)$
has bounded relative error without any assumption on $\Phi $.

First, for a chosen $\kappa \in $ $(1/n,1)$, we decompose $z_{X}(s)$ into
\begin{equation*}
z_{X}(s)=\Pr (M_{n}^{X}>s)+\Pr (S_{n}^{X}>s,\kappa s<M_{n}^{X}\leq s)+\Pr
(S_{n}^{X}>s,M_{n}^{X}\leq \kappa s).
\end{equation*}%
As for $Z_{NR1}^{X}(s)$, we use the simulation technique of \cite{4BHC13} to estimate $\Pr (S_{n}^{X}>s,\kappa s<M_{n}^{X}\leq s)$ while
$\Pr (S_{n}^{X}>s,M_{n}^{X}\leq \kappa s)$ will be estimated conditionally
on $\mathbf{W}\in \mathfrak{s}_{n}$ as for $Z_{NR2}(s)$\textbf{. }Then we
have
\begin{eqnarray*}
z_{X}(s) &=&\Pr (M_{n}^{X}>s)+\sum_{i=1}^{n}\left( \overline{F}%
_{i}(\kappa s)-\overline{F}_{i}(s)\right) \ \Pr
(S_{n}^{X}>s,X_{i}=M_{n}^{X}|\kappa s<X_{i}\leq s) \\
&&+\Pr \left( \sum_{i=1}^{n}F_{i}^{\leftarrow }\left( \Phi
(RW_{i})\right) >s,R\geq M_{n}\left\{ \frac{\Phi ^{\leftarrow }(F_{i}(\kappa
s))}{W_{i}}\right\} \right) .
\end{eqnarray*}%
Following the same rationale as for the first and second estimator, we
obtain the fourth estimator $Z_{NR4}^{X}(s)$ given by
\begin{equation*}
Z_{NR4}^{X}(s)=\Pr (M_{n}^{X}>s)+\sum_{i=1}^{n}\left( \overline{F}%
_{i}(\kappa s)-\overline{F}_{i}(s)\right) \mathbb{I}_{\{S_{n}^{X\kappa
i}>s,X_{i}^{\kappa i}=M_{n}^{X\kappa i}\}}+F_{R}(U^{X}(\mathbf{W}%
,s))-F_{R}(L_{\kappa }^{X}(\mathbf{W},s)),
\end{equation*}%
with $X_{j}^{\kappa i}=\left( X_{j}|\kappa s<X_{i}\leq s;U^{X}(\mathbf{W}%
,s)\right) $ as defined in (\ref{UX}) and
\begin{equation}
L_{\kappa }^{X}(\mathbf{W},s)=M_{n}\left\{ \frac{\Phi ^{\leftarrow
}(F_{i}(\kappa s))}{W_{i}}\right\} .  \label{LkappaX}
\end{equation}%
Similarly, for random vector $\mathbf{Y}$, we have\textbf{\ }%
\begin{eqnarray*}
z_{Y}(s) &=&\Pr (M_{n}^{Y}>s)+\sum_{i=1}^{n}\left( \bar{F}_{i}(\kappa
s)-\bar{F}_{i}(s)\right) \ \Pr (S_{n}^{Y}>s,Y_{i}=M_{n}^{Y}|\kappa
s<Y_{i}\leq s) \\
&+&\Pr \left( \sum_{i=1}^{n}\bar{F}_{i}^{\leftarrow }(\Phi
(RW_{i}))>s,R\leq M_{1}\left\{ \frac{\Phi ^{\leftarrow }(\bar{F}_{i}(\kappa
s))}{W_{i}}\right\} \right)
\end{eqnarray*}%
which leads to%
\begin{equation*}
Z_{NR4}^{Y}(s)=\Pr (M_{n}^{Y}>s)+\sum_{i=1}^{n}\left( \overline{F}%
_{i}(\kappa s)-\overline{F}_{i}(s)\right) \mathbb{I}_{\{S_{n}^{Y\kappa
i}>s,Y_{i}^{\kappa i}=M_{n}^{Y\kappa i}\}}+F_{R}(U_{\kappa }^{Y}(\mathbf{W}%
,s))-F_{R}(L^{Y}(\mathbf{W},s))
\end{equation*}%
with $Y_{j}^{\kappa i}=\left( Y_{j}|\kappa s<Y_{i}\leq s;L^{Y}(\mathbf{W}%
,s)\right) $ as defined in (\ref{LY}) and
\begin{equation}
U_{\kappa }^{Y}(\mathbf{W},s)=M_{1}\left\{ \frac{\Phi ^{\leftarrow }(%
\overline{F}_{i}(\kappa s))}{W_{i}}\right\} .  \label{UkappaY}
\end{equation}

We are able to prove that $Z_{NR4}^{Y}(s)$ is an estimator with bounded
relative error.

\begin{algorithm}
To generate a realization of $Z_{NR4}^{X}(s)$, proceed as follows:

\begin{enumerate}
\item For each $i=1,...,n$, simulate vector $(X_{1}^{\kappa
i},...,X_{n}^{\kappa i})=(X_{1},\ldots ,X_{n}|\kappa s<X_{i}\leq s)$, then
calculate $Z_{NR4,1}^{X}(s)=\sum_{i=1}^{n}\left( \overline{F}_{i}(\kappa s)-%
\overline{F}_{i}(s)\right) I_{\{S_{n}^{X\kappa i}>s,X_{i}^{\kappa
i}=M_{n}^{X\kappa i}\}}$.

\item Let $(E_{1},...,E_{n})$ be $n$ iid exponential rvs of parameter 1,
calculate $W_{i}=E_{i}/\sum_{j=1}^{n}E_{j}$.

\item Calculate $U^{X}(\mathbf{W},s)$ from (\ref{UX}) and $L_{\kappa }^{X}(%
\mathbf{W},s)$ from (\ref{LkappaX}).

\item Return $Z_{NR4}^{X}(s)=\Pr (M_{n}^{X}>s)+Z_{NR4,1}^{X}(s)+\overline{F}%
_{R}(U^{X}(\mathbf{W},s))-\overline{F}_{R}(L_{\kappa }^{X}(\mathbf{W},s))$.
\end{enumerate}
\end{algorithm}

\begin{algorithm}
To generate a realization of $Z_{NR4}^{Y}(s)$, proceed as follows:

\begin{enumerate}
\item For each $i=1,...,n$, simulate vector $(Y_{1}^{\kappa i},\ldots
,Y_{n}^{\kappa i})=(Y_{1},...,Y_{n}|\kappa s<Y_{i}\leq s)$, then calculate $%
Z_{NR4,1}^{Y}(s)=\sum_{i=1}^{n}\left( \overline{F}_{i}(\kappa s)-\overline{F}%
_{i}(s)\right) I_{\{S_{n}^{Y\kappa i}>s,Y_{i}^{\kappa i}=M_{n}^{Y\kappa
i}\}} $.

\item Let $(E_{1},...,E_{n})$ be $n$ iid exponential rvs of parameter 1,
calculate $W_{i}=E_{i}/\sum_{j=1}^{n}E_{j}$.

\item Calculate $U_{\kappa }^{Y}(\mathbf{W},s)$ from (\ref{UkappaY}) and $%
L^{Y}(\mathbf{W},s)$ from (\ref{LY}).

\item Return $Z_{NR4}^{Y}(s)=\Pr (M_{n}^{Y}>s)+Z_{NR4,1}^{Y}(s)+\overline{F}%
_{R}(U_{\kappa }^{Y}(\mathbf{W},s))-\overline{F}_{R}(L^{Y}(\mathbf{W},s))$.
\end{enumerate}
\end{algorithm}

\begin{proposition}
\label{ZNR4} $Z_{NR4}^{Y}(s)$ is an estimator with bounded relative error.
\end{proposition}

\begin{proof}
See Appendix.
\end{proof}

\section{Numerical bounds for $z_{X}(s)$ and $z_{Y}(s)$}

Inspired from the AEP algorithm in \cite{4AEP2011}, \cite{4CCMM14} have proposed sharp numerical bounds for $\Pr
(S_{n}^{X}\leq s)$\ when a closed-form expression is available for $\Pr
(X_{1}\leq x_{1},...,X_{n}\leq x_{n})$. These bounds are recalled in a first
subsection. In the next subsection, we propose an adaptation of this method
for $\Pr (S_{n}^{Y}>s)$ assuming that a closed-form expression is available
for $\Pr (Y_{1}>y_{1},...,Y_{n}>y_{n})$.

\subsection{Numerical bounds for $z_{X}\left( s\right) $}

Let us denote by $A_{S}^{\left( l,m\right) }\left( s\right) $ and $%
A_{S}^{\left( u,m\right) }\left( s\right) $ the bounds for $\Pr \left(
S_{n}^{X}\leq s\right) $\ with precision parameter $m\in \mathbb{N}^{+}$,
such that
\begin{equation*}
A_{S}^{\left( l,m\right) }\left( s\right) \leq \Pr \left( S_{n}^{X}\leq
s\right) \leq A_{S}^{\left( u,m\right) }\left( s\right) \text{, }x\geq 0%
\text{.}
\end{equation*}

Briefly, for $n=2$, $A_{S}^{\left( l,m\right) }\left( s\right) $ corresponds
to the sum of the probabilities associated to $2^{m}-1$ rectangles which lie
strictly under the diagonal $x_{1}+x_{2}=s$, i.e.%
\begin{equation}
A_{S}^{\left( l,m\right) }\left( s\right) =\sum_{i=1}^{2^{m}-1}\left(
F_{X}\left( \frac{i}{2^{m}}s,\frac{2^{m}-i}{2^{m}}s\right) -F_{X}\left(
\frac{\left( i-1\right) }{2^{m}}s,\frac{2^{m}-i}{2^{m}}s\right) \right) .
\label{chris1001}
\end{equation}%
Similarly, for $n=2$, $A_{S}^{\left( u,m\right) }\left( s\right) $ is the
sum of the probabilities associated to the $2^{m}$ rectangles strictly above
the diagonal $x_{1}+x_{2}=s$, i.e.
\begin{equation}
A_{S}^{\left( u,m\right) }\left( s\right) =\sum_{i=1}^{2^{m}}\left(
F_{X}\left( \frac{i}{2^{m}}s,\frac{2^{m}+1-i}{2^{m}}s\right) -F_{X}\left(
\frac{i-1}{2^{m}}s,\frac{2^{m}+1-i}{2^{m}}s\right) \right) .
\label{chris1002}
\end{equation}

For $n=3$, the lower bound is given by
\begin{equation}
A_{S}^{\left( l,m\right) }\left( s\right)
=\sum_{i_{1}=1}^{3^{m}-2}\sum_{i_{2}=1}^{3^{m}-1-i_{1}}\zeta _{X}^{\left(
l,m\right) }\left( s;i_{1},i_{2}\right) ,  \label{chris1003}
\end{equation}%
where%
\begin{eqnarray*}
\zeta _{X}^{\left( l,m\right) }\left( s;i_{1},i_{2}\right) &=&\Pr \left(
\frac{i_{1}-1}{3^{m}}s<X_{1}\leq \frac{i_{1}}{3^{m}}s,\frac{i_{2}-1}{3^{m}}%
s<X_{2}\leq \frac{i_{2}}{3^{m}}s,X_{3}\leq \frac{3^{m}-i_{1}-i_{2}}{3^{m}}%
s\right) \\
&=&F_{X}\left( \frac{i_{1}}{3^{m}}s,\frac{i_{2}}{3^{m}}s,\frac{%
3^{m}-i_{1}-i_{2}}{3^{m}}s\right) -F_{X}\left( \frac{i_{1}-1}{3^{m}}s,\frac{%
i_{2}}{3^{m}}s,\frac{3^{m}-i_{1}-i_{2}}{3^{m}}s\right) \\
&&-F_{X}\left( \frac{i_{1}}{3^{m}}s,\frac{i_{2}-1}{3^{m}}s,\frac{%
3^{m}-i_{1}-i_{2}}{3^{m}}s\right) +F_{X}\left( \frac{i_{1}-1}{3^{m}}s,\frac{%
i_{2}-1}{3^{m}}s,\frac{3^{m}-i_{1}-i_{2}}{3^{m}}s\right) ,
\end{eqnarray*}%
for $i_{1}=1,...,3^{m}-2$ and $i_{2}=1,...,3^{m}-1-i_{1}$. The upper bound
is given by
\begin{equation}
A_{S}^{\left( u,m\right) }\left( s\right)
=\sum_{i_{1}=1}^{3^{m}}\sum_{i_{2}=1}^{3^{m}+1-i_{1}}\zeta _{X}^{\left(
u,m\right) }\left( s;i_{1},i_{2}\right) ,  \label{chris1004}
\end{equation}%
with%
\begin{eqnarray*}
\zeta _{X}^{\left( u,m\right) }\left( s;i_{1},i_{2}\right) &=&\Pr \left(
\frac{i_{1}-1}{3^{m}}s<X_{1}\leq \frac{i_{1}}{3^{m}}s,\frac{i_{2}-1}{3^{m}}%
s<X_{2}\leq \frac{i_{2}}{3^{m}}s,X_{3}\leq \frac{3^{m}+2-i_{1}-i_{2}}{3^{m}}%
s\right) \\
&=&F_{X}\left( \frac{i_{1}}{3^{m}}s,\frac{i_{2}}{3^{m}}s,\frac{%
3^{m}+2-i_{1}-i_{2}}{3^{m}}s\right) -F_{X}\left( \frac{i_{1}-1}{3^{m}}s,%
\frac{i_{2}}{3^{m}}s,\frac{3^{m}+2-i_{1}-i_{2}}{3^{m}}s\right) \\
&&-F_{X}\left( \frac{i_{1}}{3^{m}}s,\frac{i_{2}-1}{3^{m}}s,\frac{%
3^{m}+2-i_{1}-i_{2}}{3^{m}}s\right) +F_{X}\left( \frac{i_{1}-1}{3^{m}}s,%
\frac{i_{2}-1}{3^{m}}s,\frac{3^{m}+2-i_{1}-i_{2}}{3^{m}}s\right)
\end{eqnarray*}%
for $i_{1}=1,...,3^{m}$ and $i_{2}=1,...,3^{m}+1-i_{1}$. Details for $n>3$
are provided in \cite{4CCMM14}.

\subsection{Numerical bounds for $z_{Y}\left( s\right) $}

In this section, we propose an adaptation of this method assuming that a
closed-form expression for the survival distribution function of the random
vector $\mathbf{Y}$ is available. Our objective is to develop sharp
numerical bounds, denoted $B_{S}^{\left( l,m\right) }\left( s\right) $ and $%
B_{S}^{\left( u,m\right) }\left( s\right) $, such that%
\begin{equation*}
B_{S}^{\left( l,m\right) }\left( s\right) \leq z_{Y}\left( s\right) \leq
B_{S}^{\left( u,m\right) }\left( s\right) \text{, }s\geq 0\text{.}
\end{equation*}%
Clearly, we have
\begin{equation*}
B_{S}^{\left( l,m\right) }\left( s\right) =1-A_{S}^{\left( u,m\right)
}\left( s\right) \text{ and }B_{S}^{\left( u,m\right) }\left( s\right)
=1-A_{S}^{\left( l,m\right) }\left( s\right) \text{, }s\geq 0.
\end{equation*}%
However, to achieve our goal, the task is to rewrite expressions in (\ref%
{chris1001}) and (\ref{chris1003}) for $A_{S}^{\left( u,m\right) }\left(
s\right) ,$ and (\ref{chris1002}) and (\ref{chris1004}) for $A_{S}^{\left(
u,m\right) }\left( s\right) $ such that $B_{S}^{\left( l,m\right) }\left(
s\right) $ and $B_{S}^{\left( u,m\right) }\left( s\right) $ can be defined
in terms of $\overline{F}_{Y}$. We provide expressions of the lower and
upper bounds for $n=2$ and $n=3.$

For $n=2$, we have
\begin{eqnarray*}
B_{S}^{\left( u,m\right) }\left( s\right) &=&\overline{A}_{S}^{\left(
l,m\right) }\left( s\right) =1-A_{S}^{\left( l,m\right) }\left( s\right) \\
&=&\sum_{i=1}^{2^{m}-1}\left( \overline{F}_{Y}\left( \frac{\left( i-1\right)
}{2^{m}}s,\frac{2^{m}-i}{2^{m}}s\right) -\overline{F}_{Y}\left( \frac{i}{%
2^{m}}s,\frac{2^{m}-i}{2^{m}}s\right) \right) \\
&&+\overline{F}_{Y}\left( \frac{2^{m}-1}{2^{m}}s,0\right) ,
\end{eqnarray*}%
and
\begin{eqnarray*}
B_{S}^{\left( l,m\right) }\left( s\right) &=&\overline{A}_{S}^{\left(
u,m\right) }\left( s\right) =1-A_{S}^{\left( u,m\right) }\left( s\right) \\
&=&\sum_{i=1}^{2^{m}}\left( \overline{F}_{Y}\left( \frac{i-1}{2^{m}}s,\frac{%
2^{m}+1-i}{2^{m}}s\right) -\overline{F}_{Y}\left( \frac{i}{2^{m}}s,\frac{%
2^{m}+1-i}{2^{m}}s\right) \right) +\overline{F}_{Y}\left( \frac{2^{m}}{2^{m}}%
s,0\right) .
\end{eqnarray*}

For $n=3$, we obtain%
\begin{eqnarray*}
B_{S}^{\left( u,m\right) }\left( s\right) &=&\overline{A}_{S}^{\left(
l,m\right) }\left( s\right) =1-A_{S}^{\left( l,m\right) }\left( s\right) \\
&=&\sum_{i_{1}=1}^{3^{m}-2}\sum_{i_{2}=1}^{3^{m}-1-i_{1}}\left(
\begin{array}{c}
\overline{F}_{Y}\left( \frac{i_{1}}{3^{m}}s,\frac{i_{2}}{3^{m}}s,\frac{%
3^{m}-i_{1}-i_{2}}{3^{m}}s\right) -\overline{F}_{Y}\left( \frac{i_{1}-1}{%
3^{m}}s,\frac{i_{2}}{3^{m}}s,\frac{3^{m}-i_{1}-i_{2}}{3^{m}}s\right) \\
-\overline{F}_{Y}\left( \frac{i_{1}}{3^{m}}s,\frac{i_{2}-1}{3^{m}}s,\frac{%
3^{m}-i_{1}-i_{2}}{3^{m}}s\right) +\overline{F}_{Y}\left( \frac{i_{1}-1}{%
3^{m}}s,\frac{i_{2}-1}{3^{m}}s,\frac{3^{m}-i_{1}-i_{2}}{3^{m}}s\right)%
\end{array}%
\right) \\
&&+\sum_{i_{1}=1}^{3^{m}-2}\left( \overline{F}_{Y}\left( \frac{i_{1}-1}{3^{m}%
}s,\frac{3^{m}-1-i_{1}}{3^{m}}s,0\right) -\overline{F}_{Y}\left( \frac{i_{1}%
}{3^{m}}s,\frac{3^{m}-1-i_{1}}{3^{m}}s,0\right) \right) \\
&&+\overline{F}_{Y}\left( \frac{3^{m}-2}{3^{m}}s,0,0\right) ,
\end{eqnarray*}%
and
\begin{eqnarray*}
B_{S}^{\left( l,m\right) }\left( s\right) &=&\overline{A}_{S}^{\left(
u,m\right) }\left( s\right) =1-A_{S}^{\left( u,m\right) }\left( s\right) \\
&=&\sum_{i_{1}=1}^{3^{m}}\sum_{i_{2}=1}^{3^{m}+1-i_{1}}\left(
\begin{array}{c}
\overline{F}_{Y}\left( \frac{i_{1}}{3^{m}}s,\frac{i_{2}}{3^{m}}s,\frac{%
3^{m}+2-i_{1}-i_{2}}{3^{m}}s\right) -\overline{F}_{Y}\left( \frac{i_{1}-1}{%
3^{m}}s,\frac{i_{2}}{3^{m}}s,\frac{3^{m}+2-i_{1}-i_{2}}{3^{m}}s\right) \\
-\overline{F}_{Y}\left( \frac{i_{1}}{3^{m}}s,\frac{i_{2}-1}{3^{m}}s,\frac{%
3^{m}+2-i_{1}-i_{2}}{3^{m}}s\right) +\overline{F}_{Y}\left( \frac{i_{1}-1}{%
3^{m}}s,\frac{i_{2}-1}{3^{m}}s,\frac{3^{m}+2-i_{1}-i_{2}}{3^{m}}s\right)%
\end{array}%
\right) \\
&&+\sum_{i_{1}=1}^{3^{m}}\left( \overline{F}_{Y}\left( \frac{i_{1}-1}{3^{m}}%
s,\frac{3^{m}+1-i_{1}}{3^{m}}s,0\right) -\overline{F}_{Y}\left( \frac{i_{1}}{%
3^{m}}s,\frac{3^{m}+1-i_{1}}{3^{m}}s,0\right) \right) \\
&&+\overline{F}_{Y}\left( \frac{3^{m}}{3^{m}}s,0,0\right) .
\end{eqnarray*}%
Expressions for lower and upper bounds for $n>3$ can be derived in a similar
way.

\section{Numerical study\label{section44}}

The numerical performances of the four estimators and the numerical bounds
are discussed in this section. We shall first compare both approaches by
considering small $n$ ($=2,3$) and from moderate to large $s$. We then study
the accuracy of the four estimators for the case $n=5$ where the numerical
bounds may not be computed in a reasonable time.

For both $\mathbf{X}$ and $\mathbf{Y}$, we assume that the marginal
distributions are Pareto$(\alpha _{i},1)$, i.e. $f_{X_{i}}(x)=\alpha
_{i}/(1+x)^{\alpha _{i}+1}$ for $x>0$. For the dependence structure, we
shall consider a Clayton or Gumbel copula.

The generator of the Clayton copula of parameter $\theta \in (0,\infty )$
and its inverse function are given by
\begin{equation*}
\Phi (t)=\left( 1+\frac{t}{\theta }\right) ^{-\theta }\ \text{and}\ \Phi
^{\leftarrow }(t)=\theta \left( t^{-1/\theta }-1\right) .
\end{equation*}%
The derivatives of the generator are calculated as follows
\begin{equation*}
\Phi ^{(k)}(t)=\left( 1+\frac{1}{\theta }\right) \left( 1+\frac{2}{\theta }%
\right) \ldots \left( 1+\frac{k-1}{\theta }\right) \ \left( 1+\frac{t}{%
\theta }\right) ^{-\theta -k+1}.
\end{equation*}%
The formula for the $n$-dimensional copula is
\begin{equation*}
C(u_{1},...,u_{n})=\left( u_{1}^{-1/\theta }+...+u_{n}^{-1/\theta
}-(n-1)\right) ^{-\theta }.
\end{equation*}%
Its Kendall's tau is given by $\tau =\theta ^{-1}/(2+\theta ^{-1})$. Note
that the Clayton copula has a generator satisfying the assumptions of
Proposition \ref{ZNR2}.

The generator of the Gumbel copula with parameter $b\in (0,1)$ and its
inverse function are given by
\begin{equation*}
\Phi (t)=\exp (-x^{b})\ \text{and}\ \Phi ^{\leftarrow }(t)=\left( -\log
(t)\right) ^{1/b}.
\end{equation*}%
The four derivatives of the generator are calculated as follows
\begin{eqnarray*}
\Phi ^{(1)}(t) &=&\exp (-x^{b})\left( -bt^{b-1}\right) \\
\Phi ^{(2)}(t) &=&\exp (-t^{b})\left( -b(b-1)t^{b-2}+b^{2}t^{2b-2}\right) \\
\Phi ^{(3)}(t) &=&\exp (-t^{b})\left(
-b(b-1)(b-2)t^{b-3}+3b^{2}(b-1)t^{2b-3}-b^{3}t^{3b-3}\right) \\
\Phi ^{(4)}(t) &=&\exp (-t^{b})\times \\
&&\left(
-b(b-1)(b-2)(b-3)t^{b-4}+b^{2}(b-1)(7b-11)t^{2b-4}-6b^{3}(b-1)t^{3b-4}+b^{4}t^{4b-4}\right) .
\end{eqnarray*}%
The formula for the $n$-dimensional Gumbel copula is
\begin{equation*}
C(u_{1},\ldots ,u_{n})=\exp \left( -\left[ \left( -\log (u_{1})\right)
^{1/b}+\ldots +\left( -\log (u_{n})\right) ^{1/b}\right] ^{b}\right) .
\end{equation*}%
Its Kendall's tau is given $\tau =1-b$.

\subsection{Comparison of both approaches}

\subsubsection{Numerical illustration for $z_{X}$}

In the first example, Tables \ref{Table1001} and \ref{Table1002} provide the
values of the four estimators and the numerical bounds of $z_{X}\left(
s\right) $, for $n=2$, 3, and $s=1$, $10^{2}$, $10^{4}$, and $10^{6}$. The
parameters of the Pareto distributions are given by $\alpha _{1}=0.9$, $%
\alpha _{2}=1.8$, $\alpha _{3}=2.6$, which come from Section 6 in \cite{4AEP2011} and Section 3.1 in \cite{4CCMM14}. Note
that, since the parameters of the Pareto distributions are different, the
probability $z_{X}(s)$ is equivalent to Pr$\left( X_{1}>s\right) $ for large
$s$ (e.g., Pr$\left( X_{1}>10^{6}\right) =$ 3.9811E-06).

\begin{table}[tbp] \centering%
$%
\begin{tabular}{lcccccc}
\hline\hline
$s$ & $1-A_{S}^{\left( u,20\right) }\left( s\right) $ & $1-A_{S}^{\left(
l,20\right) }\left( s\right) $ & $\mathbb{E}(Z_{NR1})$ & $\mathbb{E}%
(Z_{NR2}) $ & $\mathbb{E}(Z_{NR3})$ & $\mathbb{E}(Z_{NR4})$ \\
&  &  & $e(Z_{NR1})$ & $e(Z_{NR2})$ & $e(Z_{NR3})$ & $e(Z_{NR4})$ \\
\hline\hline
1 & \multicolumn{1}{l}{6.84165E-01} & \multicolumn{1}{l}{6.84165E-01} &
\multicolumn{1}{l}{6.83859E-01} & \multicolumn{1}{l}{6.84258E-01} &
\multicolumn{1}{l}{6.77340E-01} & \multicolumn{1}{l}{6.84340E-01} \\
& \multicolumn{1}{l}{} & \multicolumn{1}{l}{} & \textit{1.22} & \textit{1.13}
& \textit{1.18} & \textit{1.22} \\ \hline
1E02 & \multicolumn{1}{l}{1.63096E-02} & \multicolumn{1}{l}{1.63096E-02} &
\multicolumn{1}{l}{1.63378E-02} & \multicolumn{1}{l}{1.63088E-02} &
\multicolumn{1}{l}{1.63853E-02} & \multicolumn{1}{l}{1.62861E-02} \\
& \multicolumn{1}{l}{} & \multicolumn{1}{l}{} & \textit{1.99} & \textit{1.51}
& \textit{2.78} & \textit{2.45} \\ \hline
1E04 & \multicolumn{1}{l}{2.5128E-04} & \multicolumn{1}{l}{2.5128E-04} &
\multicolumn{1}{l}{2.5125E-04} & \multicolumn{1}{l}{2.5128E-04} &
\multicolumn{1}{l}{2.5129E-04} & \multicolumn{1}{l}{2.5125E-04} \\
& \multicolumn{1}{l}{} & \multicolumn{1}{l}{} & \textit{1.45} & \textit{2.04}
& \textit{2.53} & \textit{3.45} \\ \hline
1E06 & \multicolumn{1}{l}{3.9811E-06} & \multicolumn{1}{l}{3.9811E-06} &
\multicolumn{1}{l}{3.9811E-06} & \multicolumn{1}{l}{3.9811E-06} &
\multicolumn{1}{l}{3.9811E-06} & \multicolumn{1}{l}{3.9811E-06} \\
& \multicolumn{1}{l}{} & \multicolumn{1}{l}{} & \textit{1.85} & \textit{1.56}
& \textit{1.75} & \textit{1.60} \\ \hline\hline
\end{tabular}%
$%
\caption{Four estimators and numerical bounds of $z_{X}(s)$. Sum of two Pareto whose Clayton copula has Kendall's $\tau$ equal to
$\frac{3}{8}$.}\label{Table1001}%
\end{table}%

\begin{table}[tbp] \centering%
$%
\begin{tabular}{lcccccc}
\hline\hline
$s$ & $1-A_{S}^{\left( u,8\right) }\left( s\right) $ & $1-A_{S}^{\left(
l,8\right) }\left( s\right) $ & $\mathbb{E}(Z_{NR1})$ & $\mathbb{E}(Z_{NR2})$
& $\mathbb{E}(Z_{NR3})$ & $\mathbb{E}(Z_{NR4})$ \\
&  &  & $e(Z_{NR1})$ & $e(Z_{NR2})$ & $e(Z_{NR3})$ & $e(Z_{NR4})$ \\
\hline\hline
1 & \multicolumn{1}{l}{8.09108E-01} & \multicolumn{1}{l}{8.09173E-01} &
\multicolumn{1}{l}{8.08747E-01} & \multicolumn{1}{l}{8.07925E-01} &
\multicolumn{1}{l}{8.05646E-01} & \multicolumn{1}{l}{8.10322E-01} \\
& \multicolumn{1}{l}{} & \multicolumn{1}{l}{} & \textit{1.26} & \textit{1.18}
& \textit{1.19} & \textit{1.15} \\ \hline
1E02 & \multicolumn{1}{l}{1.63381E-02} & \multicolumn{1}{l}{1.63428E-02} &
\multicolumn{1}{l}{1.63361E-02} & \multicolumn{1}{l}{1.63411E-02} &
\multicolumn{1}{l}{1.63800E-02} & \multicolumn{1}{l}{1.63198E-02} \\
& \multicolumn{1}{l}{} & \multicolumn{1}{l}{} & \textit{2.14} & \textit{2.83}
& \textit{2.48} & \textit{2.53} \\ \hline
1E04 & \multicolumn{1}{l}{2.5128E-04} & \multicolumn{1}{l}{2.5128E-04} &
\multicolumn{1}{l}{2.5127E-04} & \multicolumn{1}{l}{2.5127E-04} &
\multicolumn{1}{l}{2.5129E-04} & \multicolumn{1}{l}{2.5127E-04} \\
& \multicolumn{1}{l}{} & \multicolumn{1}{l}{} & \textit{2.13} & \textit{2.51}
& \textit{2.52} & \textit{2.03} \\ \hline
1E06 & \multicolumn{1}{l}{3.9811E-06} & \multicolumn{1}{l}{3.9811E-06} &
\multicolumn{1}{l}{3.9811E-06} & \multicolumn{1}{l}{3.9811E-06} &
\multicolumn{1}{l}{3.9811E-06} & \multicolumn{1}{l}{3.9811E-06} \\
& \multicolumn{1}{l}{} & \multicolumn{1}{l}{} & \textit{1.45} & \textit{2.11}
& \textit{2.29} & \textit{1.89} \\ \hline\hline
\end{tabular}%
$%
\caption{Four estimators and numerical bounds of $z_{X}(s)$. Sum of three Pareto whose Clayton copula has Kendall's  $\tau$ equal
to $\frac{1}{6}$.}\label{Table1002}%
\end{table}%

\subsubsection{Numerical illustration for $z_{Y}$}

For the second example, the values of the four estimators and the numerical
bounds of $z_{Y}\left( s\right) $, for $n=2$, 3, and $s=1$, $10^{2}$, $%
10^{3} $, and $10^{4}$ are displayed in tables \ref{Table1003} and \ref%
{Table1004}. The parameters of the Pareto distributions are equal, with $%
\alpha _{1}=\alpha _{2}=\alpha _{3}=2.5$. In this case, all components of
the sum contribute to its large values.

\begin{table}[tbp] \centering%
$%
\begin{tabular}{lcccccc}
\hline\hline
$s$ & $B_{S}^{\left( l,20\right) }\left( s\right) $ & $B_{S}^{\left(
u,20\right) }\left( s\right) $ & $\mathbb{E}(Z_{NR1})$ & $\mathbb{E}%
(Z_{NR2}) $ & $\mathbb{E}(Z_{NR3})$ & $\mathbb{E}(Z_{NR4})$ \\
&  &  & $e(Z_{NR1})$ & $e(Z_{NR2})$ & $e(Z_{NR3})$ & $e(Z_{NR4})$ \\
\hline\hline
1 & \multicolumn{1}{l}{3.60712E-01} & \multicolumn{1}{l}{3.60712E-01} &
\multicolumn{1}{l}{3.61435E-01} & \multicolumn{1}{l}{3.61003E-01} &
\multicolumn{1}{l}{3.60885E-01} & \multicolumn{1}{l}{3.60812E-01} \\
& \multicolumn{1}{l}{} & \multicolumn{1}{l}{} & \textit{0.35} & \textit{0.14}
& \textit{0.15} & \textit{0.18} \\ \hline
1E02 & \multicolumn{1}{l}{5.14701E-05} & \multicolumn{1}{l}{5.14702E-05} &
\multicolumn{1}{l}{5.17560E-05} & \multicolumn{1}{l}{5.15248E-05} &
\multicolumn{1}{l}{5.14659E-05} & \multicolumn{1}{l}{5.13841E-05} \\
& \multicolumn{1}{l}{} & \multicolumn{1}{l}{} & \textit{0.60} & \textit{0.15}
& \textit{0.15} & \textit{0.22} \\ \hline
1E03 & \multicolumn{1}{l}{1.70171E-07} & \multicolumn{1}{l}{1.70172E-07} &
\multicolumn{1}{l}{1.71695E-07} & \multicolumn{1}{l}{1.70113E-07} &
\multicolumn{1}{l}{1.69997E-07} & \multicolumn{1}{l}{1.69823E-07} \\
& \multicolumn{1}{l}{} & \multicolumn{1}{l}{} & \textit{0.60} & \textit{0.15}
& \textit{0.15} & \textit{0.21} \\ \hline
1E04 & \multicolumn{1}{l}{5.40553E-10} & \multicolumn{1}{l}{5.40554E-10} &
\multicolumn{1}{l}{5.38901E-10} & \multicolumn{1}{l}{5.41930E-10} &
\multicolumn{1}{l}{5.40113E-10} & \multicolumn{1}{l}{5.37362E-10} \\
& \multicolumn{1}{l}{} & \multicolumn{1}{l}{} & \textit{0.61} & \textit{0.14}
& \textit{0.15} & \textit{0.21} \\ \hline\hline
\end{tabular}%
$%
\caption{Four estimators and numerical bounds of $z_{Y}(s)$. Sum of two Pareto whose Clayton survival copula has Kendall's  $\tau$
equal to $\frac{1}{2}$}\label{Table1003}%
\end{table}%

\begin{table}[tbp] \centering%
$%
\begin{tabular}{lcccccc}
\hline\hline
$s$ & $B_{S}^{\left( l,20\right) }\left( s\right) $ & $B_{S}^{\left(
u,20\right) }\left( s\right) $ & $\mathbb{E}(Z_{NR1})$ & $\mathbb{E}%
(Z_{NR2}) $ & $\mathbb{E}(Z_{NR3})$ & $\mathbb{E}(Z_{NR4})$ \\
&  &  & $e(Z_{NR1})$ & $e(Z_{NR2})$ & $e(Z_{NR3})$ & $e(Z_{NR4})$ \\
\hline\hline
1 & \multicolumn{1}{l}{4.99666E-01} & \multicolumn{1}{l}{5.00644E-01} &
\multicolumn{1}{l}{4.99080E-01} & \multicolumn{1}{l}{5.00644E-01} &
\multicolumn{1}{l}{5.00006E-01} & \multicolumn{1}{l}{5.00477E-01} \\
& \multicolumn{1}{l}{} & \multicolumn{1}{l}{} & \textit{0.47} & \textit{0.12}
& \textit{0.12} & \textit{0.21} \\ \hline
1E02 & \multicolumn{1}{l}{1.35825E-04} & \multicolumn{1}{l}{1.36732E-04} &
\multicolumn{1}{l}{1.34754E-04} & \multicolumn{1}{l}{1.36469E-04} &
\multicolumn{1}{l}{1.36570E-04} & \multicolumn{1}{l}{1.35966E-04} \\
& \multicolumn{1}{l}{} & \multicolumn{1}{l}{} & \textit{0.79} & \textit{0.13}
& \textit{0.13} & \textit{0.18} \\ \hline
1E03 & \multicolumn{1}{l}{4.58967E-07} & \multicolumn{1}{l}{4.62116E-07} &
\multicolumn{1}{l}{4.61180E-07} & \multicolumn{1}{l}{4.60001E-07} &
\multicolumn{1}{l}{4.60699E-07} & \multicolumn{1}{l}{4.60590E-07} \\
& \multicolumn{1}{l}{} & \multicolumn{1}{l}{} & \textit{0.79} & \textit{0.13}
& \textit{0.13} & \textit{0.18} \\ \hline
1E04 & \multicolumn{1}{l}{1.46118E-09} & \multicolumn{1}{l}{1.47123E-09} &
\multicolumn{1}{l}{1.47018E-09} & \multicolumn{1}{l}{1.46368E-09} &
\multicolumn{1}{l}{1.46630E-09} & \multicolumn{1}{l}{1.46335E-09} \\
& \multicolumn{1}{l}{} & \multicolumn{1}{l}{} & \textit{0.79} & \textit{0.13}
& \textit{0.13} & \textit{0.18} \\ \hline\hline
\end{tabular}%
$%
\caption{Four estimators and numerical bounds of $z_{Y}(s)$. Sum of three
Pareto whose Clayton survival copula has Kendall's $\tau$ equal to
$\frac{1}{2}$}\label{Table1004}%
\end{table}%

\subsubsection{Comments}

For both random vectors $\mathbf{X}$ and $\mathbf{Y}$, the upper and lower
bounds have been computed with the R Project for Statistical Computing.
Computation time is rather fast and varies in function of the number of
random variables $n$ and the precision parameter $m$. The evaluation of
these bounds becomes time consuming starting at $n=4$ contrarily to the
conditional Monte Carlo estimators which can be rapidly obtained no matter
the dimension $n$. For $n$ relatively small ($n=2,3$), the lower and upper
bounds are close for any value of $s$. One can see with the results of both
examples that the four conditional Monte Carlo estimators can produce values
outside of the lower and upper bounds. Both methods are complementary in the
sense that one would probably be more inclined to use the numerical bounds
in small dimension and the conditional Monte Carlo method for $n\geq 5$.

\subsection{Comparison of the four estimators when $n=5$}

We now compare the performance of our four estimators for $z_{X}$ and $z_{Y}$%
. We assume that the Pareto parameters are equal with $\alpha
_{1}=...=\alpha _{5}=2.5$. The dependence is assumed to be defined by a
Clayton copula, a Gumbel copula, a Clayton survival copula and a Gumbel
survival copula. Using Kendall's $\tau $, we study three levels of
dependence. The weak level of dependence is when $\tau =0.1$, the
intermediate level of dependence is when $\tau =0.5$ and the strong level of
dependence is when $\tau =0.9$.

For estimators $Z_{NR3}(s)$ and $Z_{NR4}(s)$, the choices of $\lambda $\ and
$\kappa $\ are sensitive. In fact, we choose the values that minimize the
numerical standard deviations of the estimators.

\begin{table}[tbp] \centering%
$%
\begin{tabular}{ccccc}
\hline\hline
Copulas & $\mathbb{E}(Z_{NR1})$ & $\mathbb{E}(Z_{NR2})$ & $\mathbb{E}%
(Z_{NR3})$ & $\mathbb{E}(Z_{NR4})$ \\[0.5ex]
& $e(Z_{NR1})$ & $e(Z_{NR2})$ & $e(Z_{NR3})$ & $e(Z_{NR4})$ \\ \hline\hline
\multicolumn{5}{c}{$s=20$, Kendall's $\tau $ = 0.1} \\ \hline\hline
Clayton & 0.00379306 & 0.00386807 & 0.00384904 & 0.00383534 \\
& \textit{2.805} & \textit{3.174} & \textit{1.916} & \textit{0.923} \\ \hline
Gumbel & 0.00734014 & 0.00722742 & 0.00711167 & 0.00718644 \\
& \textit{2.818} & \textit{1.220} & \textit{0.946} & \textit{0.666} \\ \hline
Survival Clayton & 0.00751367 & 0.00765628 & 0.00771573 & 0.007658015 \\
& \textit{2.774} & \textit{0.088} & \textit{0.095} & \textit{0.145} \\ \hline
Survival Gumbel & 0.00443284 & 0.00431637 & 0.00432776 & 0.004326611 \\
& \textit{2.916} & \textit{0.297} & \textit{0.295} & \textit{0.191} \\
\hline\hline
\multicolumn{5}{c}{$s=20$, Kendall's $\tau $ = 0.5} \\ \hline\hline
Clayton & 0.00592816 & 0.00626043 & 0.006108816 & 0.00610554 \\
& \textit{2.868} & \textit{2.132} & \textit{1.696} & \textit{0.829} \\ \hline
Gumbel & 0.01522982 & 0.01551193 & 0.015515734 & 0.015427433 \\
& \textit{2.106} & \textit{0.229} & \textit{0.229} & \textit{0.184} \\ \hline
Survival Clayton & 0.01639236 & 0.01661593 & 0.016648815 & 0.016583226 \\
& \textit{2.034} & \textit{0.112} & \textit{0.109} & \textit{0.126} \\ \hline
Survival Gumbel & 0.01095976 & 0.01116367 & 0.011170268 & 0.011168049 \\
& \textit{2.378} & \textit{0.065} & \textit{0.065} & \textit{0.110} \\
\hline\hline
\multicolumn{5}{c}{$s=20$, Kendall's $\tau $ = 0.9} \\ \hline\hline
Clayton & 0.01701347 & 0.01722898 & 0.01730004 & 0.01714553 \\
& \textit{1.909} & \textit{0.636} & \textit{0.635} & \textit{0.363} \\ \hline
Gumbel & 0.01818365 & 0.01918613 & 0.0191801 & 0.01919531 \\
& \textit{1.922} & \textit{0.028} & \textit{0.029} & \textit{0.032} \\ \hline
Survival Clayton & 0.0175496 & 0.01786598 & 0.01786844 & 0.01787014 \\
& \textit{1.966} & \textit{0.017} & \textit{0.017} & \textit{0.023} \\ \hline
Survival Gumbel & 0.01682336 & 0.01763007 & 0.01764852 & 0.01765862 \\
& \textit{1.995} & \textit{0.088} & \textit{0.089} & \textit{0.091} \\
\hline\hline
\end{tabular}%
$%
\caption{Four estimators of $z_{X}(s)$ and $z_{Y}(s)$. Sum of five Pareto
with $s=20$.}\label{Table1005}%
\end{table}%

According to the numerical results, it is remarkable that $Z_{NR1}$ has
bounded relative error. For example, under the assumption that the
dependence is a Clayton survival copula with Kendall's $\tau $ equal to $0.5$%
, when $s$ increases from $20$ (Table \ref{Table1005}) to $200$ (Table \ref%
{Table1006}), the value of $z(s)$ decreases from $0.01639236$ to $8.67011$E-$%
05$, but the relative error of $Z_{NR1}$ does not change: $2.034$ compared
to $2.036$.

Although $Z_{NR1}(s)$ is proved to have a bounded relative error under any
dependence structure, the numerical performances of this estimator is not
better than $Z_{NR2}(s)$. Note that $Z_{NR2}$ has bounded relative error
only when the dependence structure is an Archimedean survival copula of
generator $\Phi (x)=x^{-\beta }l_{\Phi }\left( x\right) $, that is the case
of Clayton survival copula in this section. However, except the case of
Clayton copula, $Z_{NR2}$ presents acceptable results in most cases. For
example, in Table \ref{Table1005} and $\tau $ = 0.5, under Gumbel survival
copula, ratio $e(Z_{NR1})/e(Z_{NR2})$ equals to $37$; or in Table \ref%
{Table1006} $\tau $ = 0.9, this ratio under Gumbel copula is approximated to
$15$.

\begin{table}[tbp] \centering%
$%
\begin{tabular}{ccccc}
\hline\hline
Copulas & $\mathbb{E}(Z_{NR1})$ & $\mathbb{E}(Z_{NR2})$ & $\mathbb{E}%
(Z_{NR3})$ & $\mathbb{E}(Z_{NR4})$ \\[0.5ex]
& $e(Z_{NR1})$ & $e(Z_{NR2})$ & $e(Z_{NR3})$ & $e(Z_{NR4})$ \\ \hline\hline
\multicolumn{5}{c}{$s=200$, Kendall's $\tau $ = 0.1} \\ \hline\hline
Clayton & 9.27623E-06 & 9.10775E-06 & 9.07197E-06 & 9.07119E-06 \\
& \textit{1.701} & \textit{3.396} & \textit{0.480} & \textit{0.270} \\ \hline
Gumbel & 3.19991E-05 & 3.16144E-05 & 3.16638E-05 & 3.13636E-05 \\
& \textit{3.216} & \textit{0.752} & \textit{0.626} & \textit{0.584} \\ \hline
Survival Clayton & 2.1843E-05 & 2.18073E-05 & 2.17775E-05 & 2.17643E-05 \\
& \textit{3.582} & \textit{0.130} & \textit{0.165} & \textit{0.145} \\ \hline
Survival Gumbel & 9.18493E-06 & 9.2342E-06 & 9.22524E-06 & 9.22872E-06 \\
& \textit{1.568} & \textit{0.130} & \textit{0.070} & \textit{0.112} \\
\hline\hline
\multicolumn{5}{c}{$s=200$, Kendall's $\tau $ = 0.5} \\ \hline\hline
Clayton & 9.54965E-06 & 9.24395E-06 & 9.37001E-06 & 9.37181E-06 \\
& \textit{2.023} & \textit{2.882} & \textit{0.182} & \textit{0.120} \\ \hline
Gumbel & 7.84723E-05 & 7.9696E-05 & 7.91973E-05 & 7.96030E-05 \\
& \textit{2.148} & \textit{0.273} & \textit{0.240} & \textit{0.230} \\ \hline
Survival Clayton & 8.67011E-05 & \textit{8.63854E-05} & 8.60928E-05 &
8.61954E-05 \\
& \textit{2.036} & \textit{0.111} & \textit{0.113} & \textit{0.133} \\ \hline
Survival Gumbel & 1.49943E-05 & 1.53712E-05 & 1.53317E-05 & 1.53699E-05 \\
& \textit{3.559} & \textit{0.263} & \textit{0.274} & \textit{0.179} \\
\hline\hline
\multicolumn{5}{c}{$s=200$, Kendall's $\tau $ = 0.9} \\ \hline\hline
Clayton & 1.0871E-05 & 1.1563E-05 & 1.03202E-05 & 1.07798E-05 \\
& \textit{2.867} & \textit{6.739} & \textit{1.077} & \textit{0.818} \\ \hline
Gumbel & 9.18427E-05 & 1.09482E-04 & 1.07552E-04 & 1.09196E-04 \\
& \textit{1.973} & \textit{0.134} & \textit{0.096} & \textit{0.127} \\ \hline
Survival Clayton & 1.14134E-04 & 9.27657E-05 & 9.27742E-05 & 9.27769E-05 \\
& \textit{1.721} & \textit{0.017} & \textit{0.017} & \textit{0.015} \\
Survival Gumbel & 7.74801E-05 & 7.93332E-05 & 7.87621E-05 & 7.94657E-05 \\
& \textit{2.155} & \textit{0.136} & \textit{0.198} & \textit{0.139} \\
\hline\hline
\end{tabular}%
$%
\caption{Four estimators of $z_{X}(s)$ and $z_{Y}(s)$. Sum of five Pareto
with $s=200$. }\label{Table1006}%
\end{table}%

The construction of $Z_{NR3}$ is more complex than that of $Z_{NR2}$;
however, the third estimator has no numerical improvement compared to the
second one except for the case of Clayton copula. Indeed, in Table \ref%
{Table1006} and $\tau $ = 0.1, the relative error of $Z_{NR3}$ is 0.480
while the relative error of $Z_{NR2}$ is 3.396 or in Table \ref{Table1006}
and $\tau $ = 0.5, the relative error of $Z_{NR3}$ is 0.182 while the
relative error of $Z_{NR2}$ is 2.882. Under the other dependence structures,
the relative errors of $Z_{NR3}$ and $Z_{NR2}$ are almost the same.

The fourth estimator has bounded relative error under Archimedean survival
copula and it presents favorable numerical results even when the dependence
structure is an Archimedean copula. For example, $Z_{NR4}$ has the smallest
relative error under Clayton copula in all tables. Under Gumbel copula,
except Table \ref{Table1005} where $s=20$ and Kendall's $\tau $ = 0.9 or
Table \ref{Table1006} where $s=200$ and Kendall's $\tau $ = 0.9, $Z_{NR4}$
also has the smallest relative error. Under Archimedean survival copulas,
there is not much difference between the relative errors of $Z_{NR2}$, $%
Z_{NR3}$ and $Z_{NR4}$.

\section{Acknowledgements}

This work was partially supported by the Natural Sciences and Engineering
Research Council of Canada (Cossette: 054993; Marceau: 053934) and by the
Chaire en actuariat de l'Universit\'{e} Laval (Cossette and Marceau:
FO502323).

\newpage

\section{Appendix\label{IntermediaryProofs4}}

\subsection{Proof of Proposition \protect\ref{Brechmann2013_fz|u1}}

From the conditional cumulative distribution function $F_{U_{j}|Z,U_{j-1},%
\ldots ,U_{1}}(u_{j}|z,u_{j-1},\ldots ,u_{1})$ with $j=1$, we derive the
conditional cumulative distribution function $F_{U_{1}|Z}$
\begin{equation*}
F_{U_{1}|Z}(u_{1}|z)=\left( 1-\frac{\Phi ^{\leftarrow }(u_{1})}{\Phi
^{\leftarrow }(z)}\right) ^{n-1}
\end{equation*}%
for $z<u_{1}<1$. Because the marginal density of $U_{1}$ is 1 on $(0,1)$,
with the density of $Z$ in Proposition \ref{Brechmann2013}, we have the
conditional density of $Z|U_{1}$
\begin{equation*}
f_{Z|U_{1}}(z|u_{1})=\frac{(\Phi ^{\leftarrow })^{(1)}(u_{1})}{(n-2)!}\left(
\Phi ^{\leftarrow }(u_{1})-\Phi ^{\leftarrow }(z)\right) ^{n-2}(\Phi
^{\leftarrow })^{(1)}(z)\Phi ^{(n)}\left( \Phi ^{\leftarrow }(z)\right)
\end{equation*}%
for $0<z<u_{1}$. The cumulative distribution function of $Z$ is obtained as
follows:%
\begin{eqnarray*}
F_{Z|U_{1}}(z|u_{1}) &=&\left( \Phi ^{\leftarrow }\right) ^{(1)}(u_{1})\frac{%
(-1)^{n-2}}{(n-2)!}\int_{0}^{z}\left( \Phi ^{\leftarrow }(v)-\Phi
^{\leftarrow }(u_{1})\right) ^{n-2}\left( \Phi ^{\leftarrow }\right)
^{(1)}(v)\Phi ^{(n)}\left( \Phi ^{\leftarrow }(v)\right) \mathrm{d}v \\
&=&-\left( \Phi ^{\leftarrow }\right) ^{(1)}(u_{1})\frac{(-1)^{n-2}}{(n-2)!}%
\int_{\Phi ^{\leftarrow }(z)}^{\infty }\left( v-\Phi ^{\leftarrow
}(u_{1})\right) ^{n-2}\Phi ^{(n)}(v)\mathrm{d}v \\
&=&-\left( \Phi ^{\leftarrow }\right) ^{(1)}(u_{1})\frac{(-1)^{n-2}}{(n-2)!}%
\int_{\Phi ^{\leftarrow }(z)}^{\infty }\left( v-\Phi ^{\leftarrow
}(u_{1})\right) ^{n-2}\mathrm{d}\left( \Phi ^{(n-1)}(v)\right) \\
&=&-\left( \Phi ^{\leftarrow }\right) ^{(1)}(u_{1})\frac{(-1)^{n-2}}{(n-2)!}%
\left( v-\Phi ^{\leftarrow }(u_{1})\right) ^{n-2}\Phi ^{(n-1)}(v)|_{\Phi
^{\leftarrow }(z)}^{\infty } \\
&&+\left( \Phi ^{\leftarrow }\right) ^{(1)}(u_{1})\frac{(-1)^{n-2}}{(n-2)!}%
\int_{\Phi ^{\leftarrow }(z)}^{\infty }(n-2)\left( v-\Phi
^{-1}(u_{1})\right) ^{n-3}\Phi ^{(n-1)}(v)\mathrm{d}v.
\end{eqnarray*}%
Note that $\lim\limits_{v\rightarrow \infty }\left( v-\Phi ^{\leftarrow
}(u_{1})\right) ^{j}\Phi ^{(j)}(v)=0$ for all $j=1,\ldots ,n-2$. The
distribution of $Z$ conditioning on $U_{1}$ is then
\begin{eqnarray*}
F_{Z|U_{1}}(z|u_{1}) &=&\left( \Phi ^{\leftarrow }\right) ^{(1)}(u_{1})\big[%
\frac{(-1)^{n-2}}{(n-2)!}\left( \Phi ^{\leftarrow }(z)-\Phi ^{\leftarrow
}(u_{1})\right) ^{n-2}\Phi ^{(n-1)}(\Phi ^{\leftarrow }(z)) \\
&-&\frac{(-1)^{n-3}}{(n-3)!}\int_{\Phi ^{\leftarrow }(z)}^{\infty }\left(
v-\Phi ^{\leftarrow }(u_{1})\right) ^{n-3}\Phi ^{(n-1)}(v)\mathrm{d}v\big] \\
&=&\ldots \ldots \\
&=&\left( \Phi ^{\leftarrow }\right) ^{(1)}(u_{1})\sum_{j=0}^{n-2}%
\frac{(-1)^{j}}{j!}\left( \Phi ^{\leftarrow }(z)-\Phi ^{\leftarrow
}(u_{1})\right) ^{j}\Phi ^{(j+1)}(\Phi ^{\leftarrow }(z))
\end{eqnarray*}%
for $z\in (0,u_{1})$.

\subsection{Proof of Proposition \protect\ref{ZNR1}}

\begin{eqnarray*}
\mathbb{V}ar(Z_{NR1}^{X}(s)) &=&\mathbb{V}ar\left( \sum_{i=1}^{n}\left(
\overline{F}_{i}(s/n)-\overline{F}_{i}(s)\right) \mathbb{I}%
_{\{S_{n}^{X_{i}}>s,X_{i}^{i}=M_{n}^{X_{i}}\}}\right) \\
&=&\sum_{i=1}^{n}\left( \overline{F}_{i}(s/n)-\overline{F}_{i}(s)\right)
^{2}Var\left( \mathbb{I}_{\{S_{n}^{X_{i}}>s,X_{i}^{i}=M_{n}^{X_{i}}\}}\right)
\\
&\leq &\sum_{i=1}^{n}\left( \overline{F}_{i}(s/n)\right) ^{2} \\
&\sim &\sum_{i=1}^{n}n^{2\alpha _{i}}\left( \overline{F}_{i}(s)\right) ^{2}
\\
&\leq &\left( \sum_{i=1}^{n}n^{2\alpha _{i}}\right) \left(
z(s)\right) ^{2}
\end{eqnarray*}%
The variance of $Z_{NR1}^{Y}(s)$ can be verified similarly.

\subsection{Proof of Proposition \protect\ref{ZNR2}}

Because $\Phi ^{(n-2)}$ is differentiable, the survival distribution
function of the radius $R$ becomes
\begin{equation*}
\overline{F}_{R}(x)=\sum_{j=0}^{n-1}(-1)^{j}\ \frac{x^{j}}{j!}\ \Phi
^{(j)}(x).
\end{equation*}%
Following the property of the regularly varying function $\Phi (x)=x^{-\beta
}l_{\Phi }\left( x\right) $, we have, for $j=1,\ldots ,(n-1)$,%
\begin{equation*}
\lim\limits_{x\rightarrow \infty }\frac{(-1)^{j}\ x^{j}\ \Phi ^{(j)}(x)}{%
\Phi (x)}=\beta (\beta +1)\ldots (\beta +j-1)
\end{equation*}%
and we can deduce
\begin{equation*}
\lim\limits_{x\rightarrow \infty }\frac{\overline{F}_{R}(x)}{\Phi (x)}%
=\lim\limits_{x\rightarrow \infty }\frac{\sum_{j=1}^{n-1}(-1)^{j}\
\frac{x^{j}}{j!}\ \Phi ^{(j)}(x)}{\Phi (x)}=\sum_{j=1}^{n-1}\frac{%
\beta (\beta +1)\ldots (\beta +j-1)}{j!}.
\end{equation*}%
We define $g(r)=\sum_{i=1}^{n}\overline{F}_{i}^{\leftharpoonup }(\Phi
(r))$ and $L_{0}^{Y}(s)=\inf \{r\in \mathcal{R}^{+}:g(r)\geq s\}$. Because $%
\overline{F}^{\leftarrow }$ and $\Phi $ are both non-increasing functions
then for all $\mathbf{W}\in \mathfrak{s}_{n}$ we have
\begin{equation*}
g(r)=\sum_{i=1}^{n}\overline{F}_{i}^{\leftarrow }(\Phi (r\times
1))\geq \sum_{i=1}^{n}\overline{F}_{i}^{\leftarrow }(\Phi (rW_{i}))
\end{equation*}%
and then $L_{0}^{Y}(s)\leq L^{Y}(\mathbf{W},s)\ $for all $\mathbf{W}\in
\mathfrak{s}_{n}$. Moreover, from the definiton of $L_{0}^{Y}(s)$, we have%
\begin{equation*}
\max\limits_{i=1,2,\ldots ,n}\overline{F}_{i}^{\leftarrow }(\Phi
(L_{0}^{Y}(s)))\geq s/n
\end{equation*}%
and
\begin{equation*}
\Phi (L_{0}^{Y}(s))\leq \max\limits_{i=1,2,\ldots ,n}\overline{F}%
_{i}(s/n)\leq n^{\alpha _{n}}\max\limits_{i=1,2,\ldots ,n}\overline{F}%
_{i}(s)\leq n^{\alpha _{n}}z(s).
\end{equation*}%
Thus, with $\lim\limits_{s\rightarrow \infty }L_{0}^{Y}(s)=\infty $, the
second moment of $Z_{NR2}^{Y}(s)$ is bounded in the following way%
\begin{eqnarray*}
\mathbb{E}\left[ \left( Z_{NR2}^{Y}(s)\right) ^{2}\right] &\leq &2\left(
\left( \Pr (M_{n}^{Y}>s)\right) ^{2}+E\left[ \left( \overline{F}_{R}(L^{Y}(%
\mathbf{W},s))^{2}\right) \right] \right) \\
&\leq &2\left( \left( \Pr (M_{n}^{Y}>s)\right) ^{2}+\left( \overline{F}%
_{R}(L_{0}^{Y}(s))\right) ^{2}\right) \\
&\sim &2\left( \left( \Pr (M_{n}^{Y}>s)\right) ^{2}+\left(
\sum_{j=1}^{n-1}\frac{\beta (\beta +1)\ldots (\beta +j-1)}{j!}\right)
^{2}\left( \Phi (L_{0}^{Y}(s))\right) ^{2}\right) \\
&\leq &2\left( \left( z(s)\right) ^{2}+\left[ \sum_{j=1}^{n-1}\frac{%
\beta (\beta +1)\ldots (\beta +j-1)}{j!}\right] ^{2}\times n^{2\alpha
_{n}}\left( z(s)\right) ^{2}\right) \\
&\leq &2\left( 1+n^{2\alpha _{n}}\left( \sum_{j=1}^{n-1}\frac{\beta
(\beta +1)\ldots (\beta +j-1)}{j!}\right) ^{2}\right) \left( z(s)\right) ^{2}
\end{eqnarray*}%
and the result follows.

\subsection{Proof of Proposition \protect\ref{ZNR3}}

We start with an inequality between $\Phi $ and $F_{R}$. \ If $\Phi $ is an $%
n$-monotone function, $(n-1)$-times differentiable and the random variable $%
R $ has the distribution function satisfies $\left( \ref{th2MN09}\right) $
then we have
\begin{equation*}
\frac{\Phi (ax)}{(1-a)^{(n-1)}}\geq \bar{F}_{R}(x),\ \forall x\in \mathbb{R}%
^{+}\mathit{\ }\text{and }a\in (0,1).
\end{equation*}%
Indeed, because $\Phi $ is non-increasing function then there exists $\mu
\in (ax,x)$ such that
\begin{equation*}
\Phi (ax)=\sum_{k=0}^{n-2}(1-a)^{k}\ \frac{x^{k}}{k!}\ (-1)^{k}\Phi
^{(k)}(x)+(1-a)^{(n-1)}\ \frac{x^{(n-1)}}{(n-1)!}\ (-1)^{(n-1)}\ \Phi ^{{%
(n-1)}}(\mu ).
\end{equation*}%
Following the property of $n$-monotone functions, $(-1)^{(n-2)}\Phi
^{(n-2)}(x)$ is a convex function, and then $(-1)^{(n-1)}\Phi ^{(n-1)}(x)$
is a non-increasing function, that means $(-1)^{(n-1)}\Phi ^{(n-1)}(\mu
)\geq (-1)^{(n-1)}\Phi ^{(n-1)}(x)$ because $\mu \leq x$. Thus we have
\begin{equation*}
\Phi (ax)\geq \sum_{k=0}^{n-1}(1-a)^{k}(-1)^{k}\ \frac{x^{k}}{k!}\
\Phi ^{(k)}(x)
\end{equation*}%
and%
\begin{equation}
\frac{\Phi (ax)}{(1-a)^{(n-1)}}\geq
\sum_{k=0}^{n-1}(1-a)^{(k-n+1)}(-1)^{k}\frac{x^{k}}{k!}\Phi
^{(k)}(x)\geq \bar{F}_{R}(x).  \label{rem33}
\end{equation}

To prove Proposition \ref{ZNR3}, first note that $Z_{NR3,2}^{Y}(s)$ is
bounded by $\bar{F}_{R}\left( L_{\lambda }^{Y}(\mathbf{W},s)\right) $ since%
\begin{equation*}
Z_{NR3,2}^{Y}(s)=F_{R}\left( U^{Y}(\mathbf{W},s)\right) -F_{R}\left(
L_{\lambda }^{Y}(\mathbf{W},s)\right) \leq \bar{F}_{R}\left( L_{\lambda
}^{Y}(\mathbf{W},s)\right) .
\end{equation*}%
Moreover, from the definition of $\overline{F}_{R}\left( L_{\lambda }^{Y}(%
\mathbf{W},s)\right) $ and $M_{2}\{\Phi ^{\leftarrow }(\bar{F}_{i}(\lambda
s))/W_{i}\}$, there exists two indexes $i_{1},i_{2}\in 1,2,\ldots ,n$ such
that
\begin{equation*}
L_{\lambda }^{Y}(\mathbf{W},s)\geq M_{2}\left\{ \frac{\Phi ^{\leftarrow }(%
\overline{F}_{i}(\lambda s))}{W_{i}}\right\} =\frac{\Phi ^{\leftarrow }(%
\overline{F}_{i_{1}}(\lambda s))}{W_{i_{1}}}\vee \frac{\Phi ^{\leftarrow }(%
\overline{F}_{i_{2}}(\lambda s))}{W_{i_{2}}}.
\end{equation*}%
Therefore,
\begin{equation*}
\left\{
\begin{array}{c@{ \geq }c}
W_{i_{1}}L_{\lambda }^{Y}(\mathbf{W},s) & \Phi ^{\leftarrow }(\overline{F}%
_{i_{1}}(\lambda s)) \\
W_{i_{2}}L_{\lambda }^{Y}(\mathbf{W},s) & \Phi ^{\leftarrow }(\overline{F}%
_{i_{2}}(\lambda s))%
\end{array}%
\right.
\end{equation*}%
implies%
\begin{equation*}
\left\{
\begin{array}{c@{ \leq }c}
\Phi \left( W_{i_{1}}L_{\lambda }^{Y}(\mathbf{W},s)\right) & \overline{F}%
_{i_{1}}(\lambda s) \\
\Phi \left( W_{i_{2}}L_{\lambda }^{Y}(\mathbf{W},s)\right) & \overline{F}%
_{i_{2}}(\lambda s)%
\end{array}%
\right. .
\end{equation*}%
Appling $\left( \ref{rem33}\right) $ with $a=W_{i_{j}}$, $j=1,2$ and $%
x=L_{\lambda }^{Y}(\mathbf{W},s)$, we have
\begin{equation*}
\left\{
\begin{array}{c@{ \leq }c}
(1-W_{i_{1}})^{n-1}\bar{F}_{R}\left( L_{\lambda }^{Y}(\mathbf{W},s)\right) &
\Phi \left( W_{i_{1}}L_{\lambda }^{Y}(\mathbf{W},s)\right) \\
(1-W_{i_{2}})^{n-1}\bar{F}_{R}\left( L_{\lambda }^{Y}(\mathbf{W},s)\right) &
\Phi \left( W_{i_{2}}L_{\lambda }^{Y}(\mathbf{W},s)\right)%
\end{array}%
\right.
\end{equation*}%
and for $j=1,2$ we have $\overline{F}_{i_{j}}(\lambda s)\sim \lambda
^{-\alpha _{i_{j}}}\overline{F}_{i_{j}}(s)\leq \lambda ^{-\alpha _{n}}z(s)$.
Finally,
\begin{eqnarray*}
\mathbb{E}\left[ \left( Z_{NR3,2}^{Y}(s)\right) ^{2}\right] \leq \mathbb{E}%
\left[ \overline{F}_{R}\left( L_{\lambda }^{Y}(\mathbf{W},s)\right) ^{2}%
\right] &\leq &\mathbb{E}\left[ \left( (1-W_{i_{1}})^{-(n-1)}\wedge
(1-W_{i_{2}})^{-(n-1)}\right) ^{2}\right] \lambda ^{-2\alpha _{n}}\left(
z(s)\right) ^{2} \\
&\leq &2^{2n-2}\lambda ^{-2\alpha _{n}}\left( z(s)\right) ^{2}.
\end{eqnarray*}

\subsection{Proof of Proposition \protect\ref{ConSurAr}}

The multivariate survival distribution function of $Y=(Y_{1},...,Y_{n})$\ is
given by
\begin{equation*}
\Pr (Y_{1}>y_{1},...,Y_{n}>Y_{n})=C\left( \overline{F}_{1}(y_{1}),...,%
\overline{F}_{n}(y_{n})\right)
\end{equation*}%
and it follows that%
\begin{equation*}
F_{Y}\left( y_{1},...,y_{n}\right) =\sum_{1\leq i_{1},\ldots
,i_{j}\leq n}(-1)^{j}\ C(\overline{F}_{i_{1}}(y_{i_{1}}),...,\overline{F}%
_{i_{j}}(y_{i_{j}})).
\end{equation*}%
We can calculate the derivative of $F_{Y}(y_{1},\ldots ,y_{n})$ following $%
\mathbf{y}_{-i}$. Note that in the sum of $2^{n}$ elements, there are only
two elements are different from $0$ after taking the derivatives $(n-1)$
times.
\begin{eqnarray*}
\frac{\partial ^{(n-1)}F_{Y}\left( y_{1},...,y_{n}\right) }{\partial
y_{1}...\partial y_{i-1}\partial y_{i+1}...\partial y_{n}} &=&\Phi
^{(n-1)}\left( \sum_{j\neq i}\Phi ^{\leftarrow }(\overline{F}%
_{j}(y_{j}))\right) \prod_{j\neq i}\left[ \left( \Phi ^{\leftarrow
}\right) ^{(1)}(\overline{F}_{j}(y_{j}))f_{j}(y_{j})\right] \\
&-&\Phi ^{(n-1)}\left( \sum_{j=1}^{n}\Phi ^{\leftarrow }(\overline{F}%
_{j}(y_{j}))\right) \prod_{j\neq i}\left[ \left( \Phi ^{\leftarrow
}\right) ^{(1)}(\overline{F}_{j}(y_{j}))f_{j}(y_{j})\right]
\end{eqnarray*}%
and note that the density of $\mathbf{Y}_{-i}$ is
\begin{equation*}
f(\mathbf{y}_{-i})=\Phi ^{(n-1)}\left( \sum_{j=1,j\neq i}^{n}\Phi
^{\leftarrow }(\overline{F}_{j}(y_{j}))\right) \prod_{j=1,j\neq i}^{n}%
\left[ \left( \Phi ^{\leftarrow }\right) ^{(1)}(\overline{F}%
_{j}(y_{j}))f_{j}(y_{j})\right] .
\end{equation*}%
The conditional distribution of $Y_{i}^{\ast }=\left( Y_{i}|\mathbf{Y}_{-i}=%
\mathbf{y}_{-i}\right) $ is then
\begin{eqnarray*}
\Pr (Y_{i}<y_{i}|\mathbf{Y}_{-i}=\mathbf{y}_{-i}) &=&1-\frac{(-1)^{n-1}\Phi
^{(n-1)}(\sum_{j=1}^{n}\Phi ^{\leftarrow }(\overline{F}%
_{j}(y_{j})))\prod_{j\neq i}\left[ \left( \Phi ^{\leftarrow }\right)
^{(1)}(\overline{F}_{j}(y_{j}))f_{j}(y_{j})\right] }{(-1)^{n-1}\Phi
^{(n-1)}(\sum_{j\neq i}\Phi ^{\leftarrow }(\overline{F}%
_{j}(y_{j})))\prod_{j\neq i}\left[ \left( \Phi ^{\leftarrow }\right)
^{(1)}(\overline{F}_{j}(y_{j}))f_{j}(y_{j})\right] } \\
&=&1-\frac{\Phi ^{(n-1)}\left( \sum_{j=1}^{n}\Phi ^{\leftarrow }(%
\overline{F}_{j}(y_{j}))\right) }{\Phi ^{(n-1)}\left( \sum_{j\neq
i}\Phi ^{\leftarrow }(\overline{F}_{j}(y_{j}))\right) }.
\end{eqnarray*}

\subsection{Proof of Proposition \protect\ref{ZNR4}}

We have%
\begin{equation*}
\Pr (S_{n}^{Y}>s,M_{n}^{Y}\leq \kappa s)=\Pr (S_{n}^{Y}>s,M_{n}^{Y}\leq
\kappa s,M_{n-1}^{Y}>\frac{1-\kappa }{n-1}\ s).
\end{equation*}%
If we estimate this probability conditionally on $\mathbf{W}\in \mathfrak{s}%
_{n}$ by the same method of estimating $Z_{NR3,2}^{Y}(s)$, the value of $%
\lambda $ in this case is $\frac{1-\kappa }{n-1}\in (0,1/n)$, the second
moment of this estimator is upper bounded by $2^{2n-2}\left( \frac{1-\kappa
}{n-1}\right) ^{-2\alpha _{n}}\times \lbrack z^{Y}(s)]^{2}$. Thus, the
variance of $Z_{NR3,2}^{Y}(s)$ is bounded by
\begin{eqnarray*}
\mathbb{V}ar(Z_{NR4}^{Y}(s)) &\leq &2\sum_{i=1}^{n}[\big(\bar{F}%
_{i}(\kappa s)-\bar{F}_{i}(s)\big)]^{2}\mathbb{V}ar\left( \mathbb{I}%
_{\{S_{n}^{Y\kappa i}>s,Y_{i}^{\kappa i}=M_{n}^{Y\kappa i}\}}\right)
+2^{2n-1}\left( \frac{1-\kappa }{n-1}\right) ^{-2\alpha _{n}}[z^{Y}(s)]^{2}
\\
&\leq &2\sum_{i=1}^{n}\left( \overline{F}_{i}(\kappa s)\right)
^{2}+2^{2n-1}\left( \frac{1-\kappa }{n-1}\right) ^{-2\alpha _{n}}\left(
z^{Y}(s)\right) ^{2} \\
&\leq &\left( 2\kappa ^{-2\alpha _{n}}+2^{2n-1}\left( \frac{1-\kappa }{n-1}%
\right) ^{-2\alpha _{n}}\right) \left( z^{Y}(s)\right) ^{2}.
\end{eqnarray*}

\newpage

\bibliographystyle{apalike}
\bibliography{bilbio}

\end{document}